\documentclass[10pt, final, conference, letterpaper, onecolumn, oneside]{IEEEtran}

\usepackage{cite}
\usepackage{graphicx}
\usepackage{amsmath}
\usepackage{amssymb}
\usepackage{mathtools}
\usepackage{bm}
\usepackage{epsfig}
\usepackage{times}
\usepackage{color}
\usepackage{url}
\usepackage{array}
\usepackage{multirow}
\usepackage{rotating}
\usepackage{extarrows}
\usepackage{mathabx} %for producing double, triple primes
\usepackage{wasysym}
\usepackage{ulem} %package to underline (\uline, \uuline, \uwave), strike-through (\sout), and masking the text (\xout)
\usepackage{braket}
\usepackage{pifont}
\usepackage{soul}
\usepackage{enumitem}
\usepackage{float}
\usepackage{tikz}
\usepackage{titlesec}
\usepackage{booktabs} % For better horizontal rules
\usepackage{subcaption}

\usetikzlibrary{positioning}
\titleclass{\subsubsubsection}{straight}[\subsection]
\newcounter{subsubsubsection}[subsubsection]
\renewcommand{\thesubsubsubsection}{\thesubsubsection.\arabic{subsubsubsection}}
\titleformat{\subsubsubsection}
{\normalfont\normalsize\bfseries}{\thesubsubsubsection}{1em}{}
\titlespacing*{\subsubsubsection}
{0pt}{3.25ex plus 1ex minus .2ex}{1.5ex plus .2ex}

\newcolumntype{P}[1]{>{\centering\arraybackslash}p{#1}} %for centering column values in a table

\begin{document}
\title{Lossless Image Compression Using Multi-level Dictionaries: Binary Images}%
\author{\IEEEauthorblockN{Samar Agnihotri, Renu Rameshan, and Ritwik Ghosal}\\%
Correspondence Email: samar.agnihotri@gmail.com%
}

\maketitle
\begin{abstract}
Lossless image compression is required in various applications to reduce storage or transmission costs of images, while requiring the reconstructed images to have zero information loss compared to the original. Existing lossless image compression methods either have simple design but poor compression performance, or complex design, better performance, but with no performance guarantees. In our endeavor to develop a lossless image compression method with low complexity and guaranteed performance, we argue that compressibility of a color image is essentially derived from the patterns in its spatial structure, intensity variations, and color variations. Thus, we divide the overall design of a lossless image compression scheme into three parts that exploit corresponding redundancies. We further argue that the binarized version of an image captures its fundamental spatial structure. In this first part of our work, we propose a scheme for lossless compression of binary images.

The proposed scheme first learns dictionaries of $16\times16$, $8\times8$, $4\times4$, and $2\times 2$ square pixel patterns from various datasets of binary images. It then uses these dictionaries to encode binary images. These dictionaries have various interesting properties that are further exploited to construct an efficient and scalable scheme. Our preliminary results show that the proposed scheme consistently outperforms existing conventional and learning based lossless compression approaches, and provides, on average, as much as $1.5\times$ better performance than a common general purpose lossless compression scheme (WebP), more than $3\times$ better performance than a state of the art learning based scheme, and better performance than a specialized scheme for binary image compression (JBIG2).
\end{abstract}

\section{Introduction}
\label{sec:intro}
Lossless compression of images strives to exploit various redundancies in images to reduce their sizes with the constraint that the reconstructed image has the same quality and information content as the original. Applications in various domains such as medical, forensics, document analysis, data archival may require relevant images to be losslessly compressed to reduce storage and transmission costs without any loss of information. In the past few decades numerous schemes and commercial standards have been developed towards this end, such as GIF \cite{gif89a}, PNG \cite{PNG}, TIFF \cite{tiff}, JPEG-LS \cite{jpegls}, WebP \cite{googleCT}, and JPEG-XL \cite{jpegxl} and are used billions of times everyday to store and transmit images.

All existing schemes for lossless image compression are broadly of three types: universal source coding based, prediction based, and distribution learning based. Schemes based on universal source coding, such as \cite{096storer, 097storerHelfgott, 101rizzoStorerCarpentieri, 106wuLonardiSzpankowski, 108baeJuang}, use variants of Lempel-Ziv coding and linearize an image in rows or strips of pixels or create a ``front'' of pixels which are already explored and then use some universal source coding scheme to compress those losslessly. Prediction based schemes such as \cite{jpegls, googleCT, jpegxl}, on the other hand divide an image in various segments based on some complexity heuristic, and then use prediction and entropy coding over the prediction error. Finally, the distribution learning based schemes, such as \cite{mentzer}, attempt to estimate the probability distribution of pixel values or their residues using self-supervised learning.

Impressive as many of these schemes are for their usage and performance, all these schemes are based on heuristics, drawing upon some general theoretical principles and empirical results. However, many that perform well are too complex for rigorous performance analysis and those that are simple enough for analysis do not perform too well. Further, given their heuristic nature, none of these schemes consistently perform better than all other schemes over all types of images and no guarantees of their respective performance can be given.

In our pursuit of developing a lossless image compression scheme that performs better than all other schemes over all types of images, is amenable for rigorous performance analysis, and can provided guaranteed compression for a given type of images, we step back and ask the most basic question related to image compression: where does the information lie in an image? In other words, in an image where do the patterns, redundancies, or correlations lie that can be exploited to reduce the size of the losslessly compressed image?

The reader may recall that in the RGB color space, the color of each image pixel is encoded using $24$ bits, with $8$ bits used to represent the intensities of each of the R, G, and B components. While a pixel in a color image is represented using $24$ bits, in a grayscale image it uses $8$ bits, and in a binary or bilevel image it uses only two bits. In this work, we hypothesize that a RGB image has information in its spatial structure, and intensity and colors variations with core information captured by its spatial structure and then intensity and colors variations adding further layers of information in that order. We argue that the primary spatial structure of an image is captured by its binary version\footnote{It should be noted that further spatial structure may appear in a binary image when intensity and colors are added to it. However, such secondary structures can be addressed by considering patterns in intensities and colors of the neighboring pixels.} and the primary intensity variations in an image are captured by its grayscale version\footnote{Patterns in intensity variations in a grayscale image may appear when color is added to them and those can be exploited using the correlation among their color values.}. Therefore, overall compressibility of a RGB image can be considered as a function of its compressibilities with respect to its spatial structure, and its intensity and color variations. In other words, the overall compressibility of a color image can be understood and explained in terms of the compressibilities of its binary and grayscale versions and color variations.

In this work, as the first part of our effort to develop a lossless color image compression scheme, we discuss the design of a lossless binary image compression scheme. We strive to provide an introduction to this new approach for lossless binary image compression and report its performance compared to existing conventional and learning based approaches. However, we assert that the current implementation of the proposed scheme only provides a proof of the concept. There exist various opportunities to optimize it to improve its compression performance as well as computational requirements. We plan to exploit such opportunities in our immediate work. Still, it is encouraging to note that even in its not fully optimized form, the proposed scheme outperforms all existing commercial and/or state-of-the-art schemes for lossless image compression and provides competitive performance with respect to a specialized scheme for binary image compression. This points to significant potential of such schemes for the lossless image compression.

The proposed scheme to encode a binary image includes two major steps: multi-level dictionary learning and image encoding. We propose to cover an image with square pixel blocks of sizes $16\times16$, $8\times8$, $4\times4$, and $2\times 2$. Therefore, we first create dictionaries of patterns for each of these block-sizes from given datasets of binary images and then use these dictionaries to encode a binary image. We provide the details in Section~\ref{sec:proposal}.

We plan to introduce schemes for lossless compression of grayscale and color images in our upcoming work. Though those schemes follow the same general approach as the scheme introduced in this work for binary images, yet those introduce major innovations to manage the computational costs while providing better compression performance than the existing schemes.

\subsection{Contributions}
\label{subsec:contri}
\begin{itemize}
\item We introduce a new paradigm for the lossless compression of color images where the overall compression scheme is developed in a layered manner by successively exploiting the patterns in spatial structure, and intensity and color variations.
\item We argue that the fundamental spatial structure of an image is captured by its binarized version.
\item We provide a multi-level dictionary learning based approach to losslessly compress and decompress binary images.
\item Our results establish that the proposed scheme consistently outperform major conventional and learning based lossless compression schemes and provides, on average, as much as $1.5\times$ better performance than a common general purpose lossless compression scheme (WebP), more than $3\times$ better performance than a state of the art learning based scheme \cite{mentzer}, and better performance than a specialized scheme for binary image compression (JBIG2).
\end{itemize}

\subsection{Related Work}
\label{subsec:relatedWork}
Though we approach lossless compression of binary images as a first step in developing an effective scheme for lossless compression of grayscale and color images,  in literature lossless compression of binary images has been widely addressed for its own importance in processing and storing text, maps, seismograms, fingerprints, logos, silhouettes, line drawings, facsimile etc. Accordingly a lot of work exists for lossless compression of binary images. A facsimile standard, called JBIG2 \cite{100jbig2}, segments an image into different regions and uses different coding schemes to encode them. However, this requires a good segmentation algorithm. A scheme for lossless compression of binary and grayscale images, called Block Arithmetic Coder for Image Compression (BACIC) \cite{101reavyBoncelet}, uses variable-to-fixed block arithmetic coder to encode symbols whose probability distribution is estimated using a single context model throughout the whole image. However, this may result in poor compression for images which may have regions with widely different context, such as text and halftone. The authors in \cite{106xiaoBoncelet} propose a scheme that adaptively weighs different context models based on their relative accuracy in real-time, where the weighting factors are updated in a pixel-by-pixel manner using the relative accuracy of the corresponding context models. However, this scheme in spite of its higher complexity achieves only a marginal performance gain with respect to simpler BACIC. In \cite{101imJeong}, a lossless compression algorithm is proposed that creates a context for each pixel by rearranging a 2-D image into a corresponding 1-D string and encodes it using hierarchical enumerative coding. However, its compression performance depends on scanning order of the 2-D images to create the corresponding 1-D strings and this may lead to substantially poor compression performance as it may create spurious long-range correlations by putting nearby pixels far away.

\subsection{Organization}
\label{subsec:organization}
The paper is organized as follows. We provide the details of the proposed method, such as dictionary learning and encoder design in Section~\ref{sec:proposal}. In Section~\ref{sec:results}, we provide a description of datasets used for dictionary learning and results comparing the compression ratios of the proposed scheme with those of various other major schemes. Section~\ref{sec:bckgrnd} discusses statistical analysis of images and learned multi-level dictionaries that guided the design of the proposed approach and provides insights into the workings of the proposed approach and its performance. Finally, Section~\ref{sec:concl} concludes the paper and provides some directions for further work.

\section{The Proposed Scheme}
\label{sec:proposal}
In this work, we take an approach that is very different from context-modeling approach adopted by most of the existing work on lossless compression of binary images. One of the major results of \cite{103feldmanCrutchfield} is that the 2-D entropy density of an infinite 2-D lattice monotonically decreases to a limit as the side-lengths of rectangular blocks covering the lattice tend to infinity, while keeping the ratio of the side-lengths finite. Motivated by this, we propose to cover an image with non-overlapping pixel squares of side-length $2^m, m \ge 1$. Then, each of these squares is replaced by its corresponding entropy code obtained from the dictionary of all $2^m \times 2^m$ patterns with their respective frequencies, learned \textit{a priori} over a given collection of binary images. Next, we provide the details of multi-level dictionary learning and encoder design.

\subsection{Multi-level dictionary learning}
\label{subsec:diclearn}
We use multi-level dictionaries to compress a given image.  Dictionaries contain string representations of image patches and their frequencies. As our focus is on binary images, the patches are represented using hex strings. Multi-level refers to the four patch sizes that are being used. To capture the statistics at a very small scale, we use $2\times2$ patches and we capture the statistics at increasingly higher levels by considering square patches of sizes $4,\,8,$ and $16$. 

The statistics of $n \times n$ patches, where $n = 2,\,4,\, 8,\, 16$ are learned from a wide variety of images, derived from various publicly available datasets. While for patch sizes $2$ and $4$, the number of patches are small enough to be handled without any special considerations in coding, the case for sizes $8$ and $16$ are not so. Despite the combinatorial explosion, we processed several giga bytes of data and have come up with interesting observations and exploited those to design an efficient encoder. Specific details of the datasets used, processing steps, and corresponding observations are detailed in Section~\ref{sec:results}.

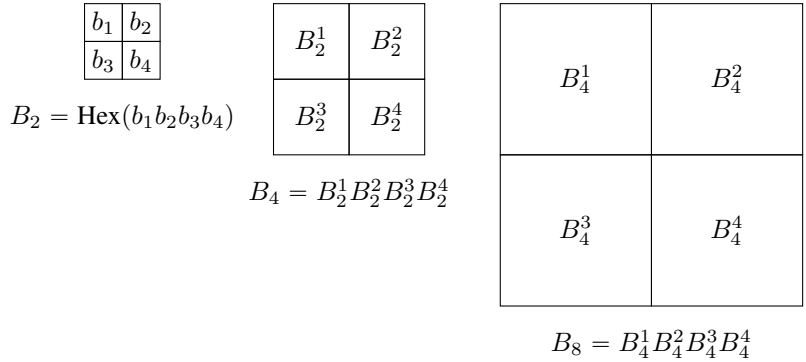
\begin{figure}[H]
  \centering
  \begin{tikzpicture}[cell/.style={minimum size=0.5cm, draw, align=center, inner sep=0, outer sep=0},
    bigcell/.style={minimum size=1cm, draw, align=center, inner sep=0, outer sep=0},
    biggcell/.style={minimum size=2cm, draw, align=center, inner sep=0, outer sep=0},
    ]

    % First square with four cells
    \node[cell, anchor=north west] (b1) at (0,2) {$b_1$};
    \node[cell, right=0cm of b1] (b2) {$b_2$};
    \node[cell, below=0cm of b1] (b3) {$b_3$};
    \node[cell, right=0cm of b3] (b4) {$b_4$};

    % Text below the first square
    \node[below=0.5cm of b3.south east, anchor=center] {$B_2 = \text{Hex}(b_1b_2b_3b_4)$};

    % Second square to the right with sides twice the size
    \node[bigcell, anchor=north west, at=(b2.north east)] (B1) at (2.5,2) {$B_2^1$};
    \node[bigcell, right=0cm of B1] (B2) {$B_2^2$};
    \node[bigcell, below=0cm of B1] (B3) {$B_2^3$};
    \node[bigcell, right=0cm of B3] (B4) {$B_2^4$};

    % Text below the second square
    \node[below=0.5cm of B3.south east, anchor=center] {$B_4 = B_2^1B_2^2B_2^3B_2^4$};

    % Third square to the right with sides twice the size of second
    \node[biggcell, anchor=north west, at=(b2.north east)] (BB1) at (5.5,2) {$B_4^1$};
    \node[biggcell, right=0cm of BB1] (BB2) {$B_4^2$};
    \node[biggcell, below=0cm of BB1] (BB3) {$B_4^3$};
    \node[biggcell, right=0cm of BB3] (BB4) {$B_4^4$};

    % Text below the second square
    \node[below=0.5cm of BB3.south east, anchor=center] {$B_8 = B_4^1B_4^2B_4^3B_4^4$};

  \end{tikzpicture}
  \caption{The bits $b_i$ or blocks $B_n^i$ are read in a raster scan order; here $n$ indicates the patch size and $i$, the bit or the block.}
  \label{fig:1}
\end{figure}

As a first step in creating the dictionary, the RGB input images are binarized using Otsu's thresholding \cite{otsu}. Dictionaries are created separately for each patch size. For each value of $n$, all possible patches of size $n \times n$ are read from an image, binarized, converted to a hex string, and are stored in an array. The order in which bits are read from a patch to form the hex string is shown in Figure~\ref{fig:1}. For $2\times2$ patches the bits are read out in a zig-zag manner. A $4 \times 4 $ patch is formed by arranging four $2 \times 2$ patches, again in a zig-zag manner. This process is repeated for larger size patches as illustrated in Figure~\ref{fig:1}.

Patches are extracted from images and stored in an array. To ensure that patches are picked in an independent manner for counting their respective frequencies, the array of image patches is shuffled across an image, across images in a dataset, and across datasets.

\subsection{Multi-level image encoding}
\label{subsec:encoding}
Having constructed dictionaries of all unique $16\times16$, $8\times8$, $4\times4$, and $2\times2$ patches that occur in the training data along with their respective frequencies, we next use these dictionaries to compress a given image as follows. To keep the exposition simple, here we discuss the encoding only for the images with side-lengths that are a multiple of $16$, though the description can be easily extended to images with arbitrary side-lengths.

To compress a given image, we start with the largest available patch size, which is $16\times16$ in this work. We divide the entire image in terms of $16\times16$ non-overlapping patches. Among all these patches, we find the ones which occur in the corresponding $16\times16$ dictionary. Using frequencies of those patches, we construct the corresponding Canonical Huffman codes (CHCs) and replace these patches in the image by the corresponding CHCs. Each of those $16\times16$ patches in the image which are not found in the $16\times16$ dictionary are divided into four $8\times8$ patches. Among all such $8\times 8$ patches, we find the ones which occur in the corresponding $8\times8$ dictionary, construct CHCs for those, and replace those patches in the image by the corresponding CHCs. Further, each of those $8\times8$ patches in the image which are not found in the $8\times8$ dictionary are divided into four $4\times4$ patches and the same process as for $8\times8$ patches is repeated, but now for $4\times4$ patches. Lastly, the same goes for $2\times2$ patches.

\begin{figure}[!t]
  \centering
  \includegraphics[scale=0.4]{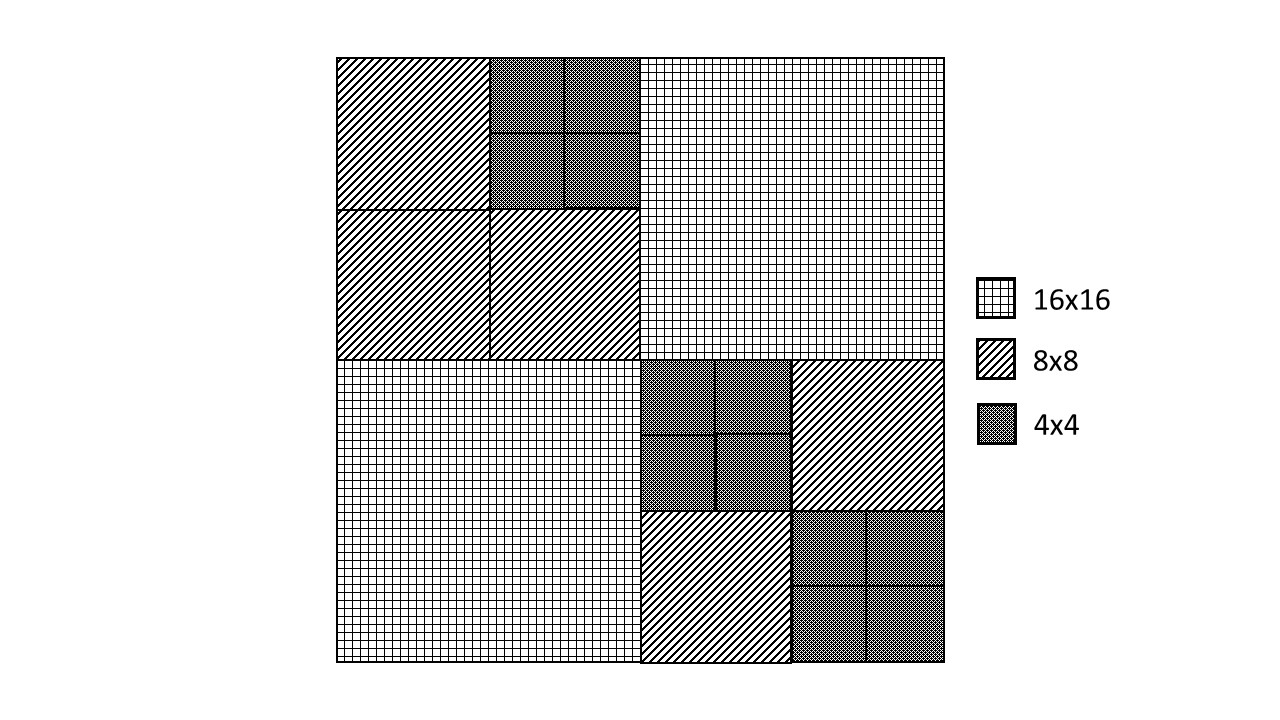}
  \caption{Illustration of hybrid encoding for a $32 \times 32$ patch.}
  \label{fig:enc}
\end{figure}

In Figure~\ref{fig:enc}, we provide an example of compressing a $32 \times 32$ image successively with smaller block sizes, starting with the block size of $16 \times 16$.

A small header is prefixed to the compressed image to help the decoder know the original image size, identities of patterns from various dictionaries for which the CHCs are created and which are subsequently used to encode the corresponding patches, and the identities of the patches in the original image so encoded. However, we skip the details of this header to keep the discussion simple.

For decompression, we follow the reverse process of compression and recreate the original image by referring to the dictionaries, which are shared \textit{a priori} with the decoder, to find the patterns corresponding to codes in the compressed image and use those to replace the codes in the compressed image to reconstruct the original image.

\section{Results}
\label{sec:results}
We compare the proposed method with widely used lossless compression algorithms both from engineered solution category like PNG \cite{PNG}, WebP \cite{googleCT}, JPEG-XL \cite{jpegxl}, JBIG2 \cite{100jbig2}, and also a deep learning based model \cite{mentzer}. Binary images are given as input to all the algorithms. 

\subsection{Deep learning based model}
As we are proposing a learning based method, we compare the performance of the proposed scheme with that of a recent deep learning based lossless compression scheme \cite{mentzer}. This method uses a convolutional neural network along with the lossy compression scheme BPG \cite{bpg} for achieving lossless compression. BPG itself is based on the HEVC video coding standard \cite{HEVC}.

In \cite{mentzer}, the CNN model learns the conditional distribution of natural images, given as side information the residual image obtained as the difference of the actual image and the image reconstructed from the BPG format. Two sub-networks are learned: a Q-Classifier network which provides the quantization parameter to BPG and a residual compressor, which predicts the conditional probabilities ($p$). Once $p$ is learned, an arithmetic encoder uses this to generate the bit stream.

We retrain the network using binary images. As in \cite{mentzer}, for training we use the Open Images \cite{openimage} dataset. This requires around $190$ Gb of data.

\subsection{Datasets}
\label{subsec:datasets}
For dictionary learning, we use various datasets from ImageNet \cite{deng2009imagenet} to the recent SAM \cite{sam} dataset. While some of them are class specific, SAM brings in variety by having images across several domains.  

The class specific datasets that we use are ImageNet \cite{deng2009imagenet}, 256 Object Categories \cite{256OC}, iNaturalist7 \cite{iNat}, and DTD \cite{dtd}. The first two contain images for object categorization, the third one contains fine-grained images of plants and animals; and DTD has different classes of textural images. As mentioned before, to avoid any bias in the way patches are sampled from images, we create new sets of images by randomly sampling images from across classes, and across datasets. We restructure the above datasets and create $732$ directories each with $1132$ images. This amounts to $88$ GB of data. Let us call this combined dataset as D1 to distinguish it from the SAM dataset described next. 

From the SAM training data we use $50$ directories. We do not mix the images in this case as SAM being a segmentation dataset, the images are diverse within and across directories. Each directory has around $11$ GB of data. Sample images from these datasets are shown in Figure~\ref{fig:sample}.

For $8 \times 8$ and $16 \times 16$ patches, as the total number of possible patches are huge, $2^{64}$ and $2^{256}$, respectively, the number of observed patches are also quite high despite the fact that many patches repeat with a high frequency. We processed roughly 638 GB of data which amounts to a huge number of patches. A large fraction of these patches are quite rare. In Section~\ref{sec:bckgrnd}, we provide more insights into the observed statistics of these learned dictionaries. We process the unit frequency patches and patches with frequency greater than one, separately. This separation is maintained for all type of processing for patches of size $8$ and $16$.  

\subsection{Compression ratio comparison}
\label{subsec:CRcomparison}
\begin{figure}[H]
  \centering
  
  \begin{subfigure}[b]{0.23\textwidth}
    \includegraphics[width=\textwidth]{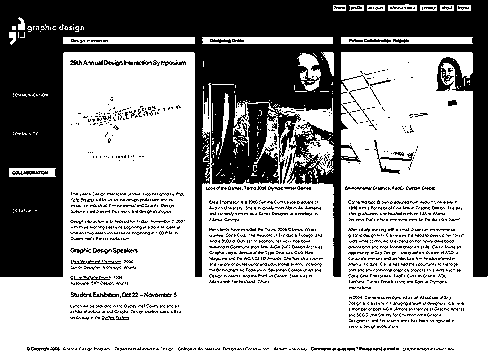}
    \caption{ n06359193}
    \label{fig:d1}
  \end{subfigure}
  \begin{subfigure}[b]{0.23\textwidth}
    \includegraphics[width=\textwidth,height=0.748\textwidth]{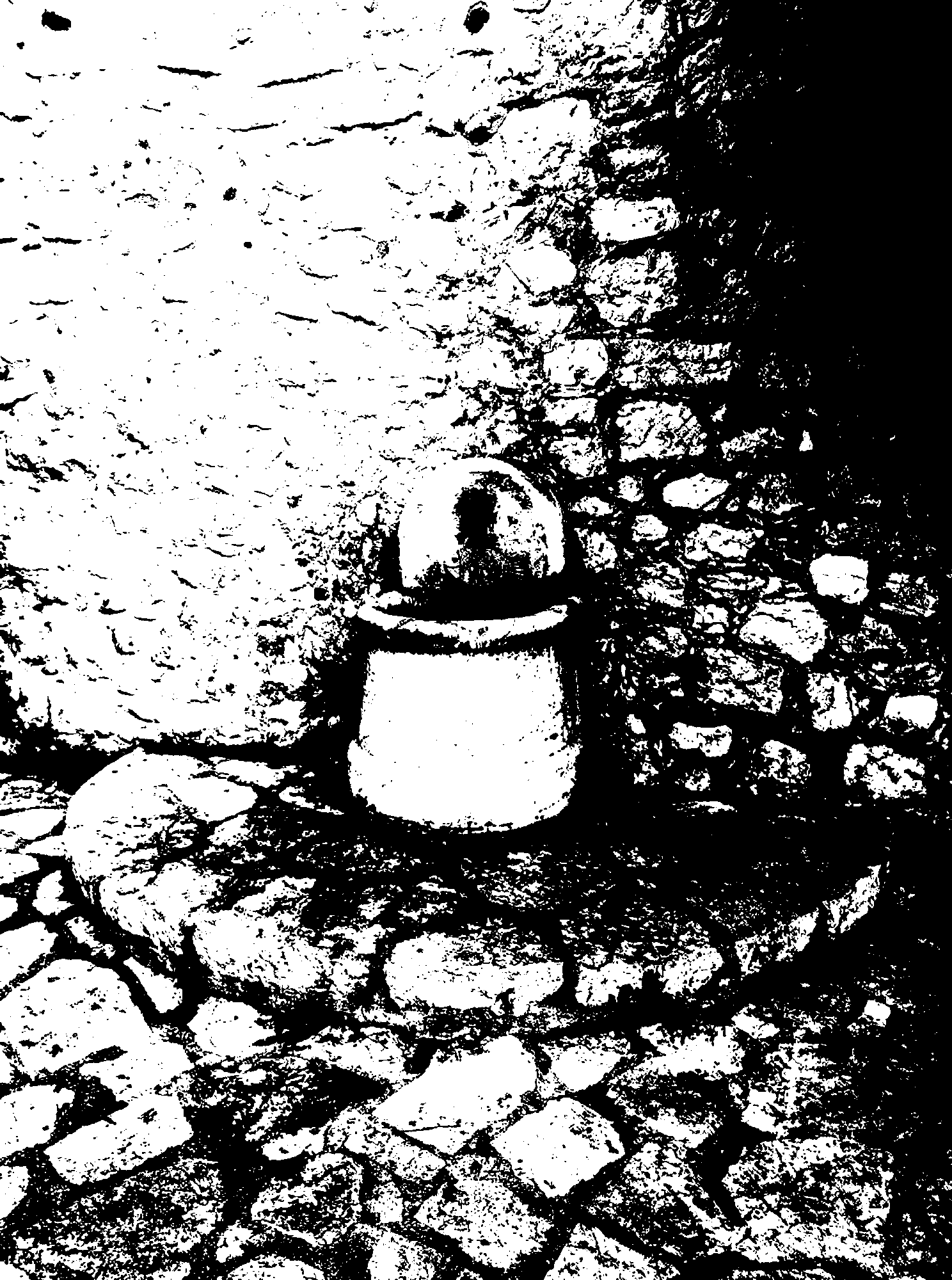}
    \caption{ sa\_6369140}
    \label{fig:d2}
  \end{subfigure}
  \begin{subfigure}[b]{0.23\textwidth}
    \includegraphics[width=\textwidth]{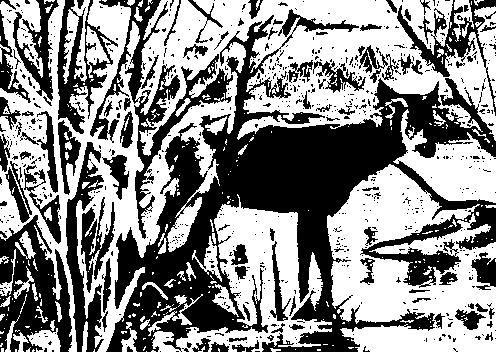}
    \caption{ 4b474218}
    \label{fig:d3}
  \end{subfigure}
  \begin{subfigure}[b]{0.23\textwidth}
    \includegraphics[width=\textwidth]{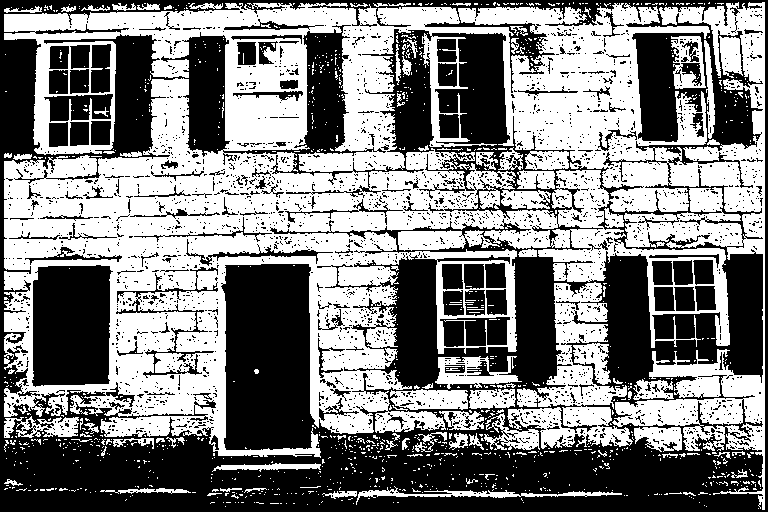}
    \caption{ kodim01}
    \label{fig:d4}
  \end{subfigure}
  \caption{Sample images from each dataset. Left to right, ImageNet \cite{deng2009imagenet}, SAM \cite{sam}, iNaturalist \cite{iNat}, and Kodak \cite{kodak}.
}
\label{fig:sample}
\end{figure}

We compare the performance of the proposed method to engineered solutions like PNG \cite{PNG}, WebP \cite{googleCT}, JPEG-XL \cite{jpegxl}, and JBIG2 \cite{100jbig2}, and a learning based solution MGT \cite{mentzer} with respect to sample images extracted from four different datatsets: ImageNet \cite{deng2009imagenet}, SAM \cite{sam}, iNaturalist \cite{iNat}, and Kodak \cite{kodak}. The Kodak image dataset contains high-resolution true-color images for compression testing. Binarized representative test images from each of these datasets are illustrated in Figure~\ref{fig:sample}.

The compression ratios of the proposed scheme and the comparison pool are shown in Table~\ref{tab:table1}. In the appendix, we provide comparison results over more images from each of these four datasets. It can be observed from these results that the proposed scheme consistently outperforms the general-purpose compressors, such as PNG and JPEG-XL, by a large margin. Also, it outperforms WebP by $1.5\times$, on average, though there are instances where the proposed scheme is only slightly better than WebP and for a handful of cases, WebP outperforms the proposed scheme by a small margin.

It can be observed that the proposed scheme comprehensive outperforms a state of the art learning based scheme \cite{mentzer} and provides, on average, more than $3\times$ better performance than it.

Further to address a potential concern that it may be unfair to compare the performance of the proposed scheme with schemes which are not optimized for lossless compression of binary images, we also compare the performance of the proposed scheme with that of JBIG2 \cite{100jbig2} - a specialized scheme for binary image compression. It can be seen from results in various tables that though JBIG2 performs better than the proposed scheme for some binary images, the proposed scheme, even in its yet-to-be-fully-optimized form, still performs better than it, on average, across all datasets.

Lastly, to address another potential concern that it may not be fair to compare the performance of the proposed scheme with that of JBIG2 over the binarized versions of general images when JBIG2 is meant to compress images corresponding to scanned documents and facsimile, we compare the performance of the proposed scheme with that of JBIG2 over images from FigureQA \cite{figureQA} - a dataset of plots, graphs, and charts; and scanned document binary images from various sources on the Internet, such as \cite{ITU-T.T.24}. Our results show that the two schemes still have comparable performance, even when JBIG2 is highly optimized for such images and the proposed scheme is yet to be fully optimized for such images as it uses dictionaries which are not trained over such images, but over binarized images derived from datasets with very different nature, as detailed in subsection~\ref{subsec:datasets}.

\begin{table}[!t]
  \centering
  \begin{tabular}{|P{1.5cm}|P{1.5cm}|P{1.5cm}|P{1.5cm}|P{1.5cm}|P{1.5cm}|P{1.5cm}|}
    \hline
    \multicolumn{1}{|c|}{Image} & \multicolumn{1}{c|}{Proposed} & \multicolumn{1}{c|}{MGT} & \multicolumn{1}{c|}{PNG} & \multicolumn{1}{c|}{WebP} & \multicolumn{1}{c|}{JPEG-XL} &  \multicolumn{1}{c|}{JBIG2} \\
    \hline
    \multicolumn{1}{|c}{} & \multicolumn{5}{c}{ImageNet} & \multicolumn{1}{c|}{} \\
    \hline
    \multirow{1}{*}{n02910353} & \multirow{1}{*}{\textbf{121.2}} & \multirow{1}{*}{30.71} & \multirow{1}{*}{32.45} & \multirow{1}{*}{52.92} & \multirow{1}{*}{6.24} & \multirow{1}{*}{86.48} \\
    % & & & & & &\\
    \hline
    \multirow{1}{*}{n03196217} & \multirow{1}{*}{\textbf{332.24}} & \multirow{1}{*}{93.56} & \multirow{1}{*}{62.93} & \multirow{1}{*}{197.47} & \multirow{1}{*}{18.88} & \multirow{1}{*}{298.14}\\
    % & & & & & &\\
    \hline
    \multirow{1}{*}{n06359193} & \multirow{1}{*}{31.55} & \multirow{1}{*}{10.51} & \multirow{1}{*}{15.72} & \multirow{1}{*}{31.68} & \multirow{1}{*}{3.82} & \multirow{1}{*}{\textbf{35.25}}\\
    % & & & & & &\\
    \hline
    \multicolumn{1}{|c}{} & \multicolumn{5}{c}{SAM} & \multicolumn{1}{c|}{} \\
    \hline
    \multirow{1}{*}{sa\_6367649} & \multirow{1}{*}{40.52} & \multirow{1}{*}{13.887} & \multirow{1}{*}{16.76} & \multirow{1}{*}{32.27} & \multirow{1}{*}{4.69} & \multirow{1}{*}{\textbf{45.76}}\\
    % & & & & & &\\
    \hline
    \multirow{1}{*}{sa\_6367685} & \multirow{1}{*}{147.95} & \multirow{1}{*}{47.01} & \multirow{1}{*}{36.86} & \multirow{1}{*}{100.61} & \multirow{1}{*}{10.14}& \multirow{1}{*}{\textbf{168.2}} \\
    % & & & & & &\\
    \hline
    \multirow{1}{*}{sa\_6369140} & \multirow{1}{*}{45.06} & \multirow{1}{*}{13.33} & \multirow{1}{*}{18.56} & \multirow{1}{*}{32.19} & \multirow{1}{*}{4.45}& \multirow{1}{*}{\textbf{51.69}} \\
    % & & & & & &\\
    \hline
    \multicolumn{1}{|c}{} & \multicolumn{5}{c}{iNaturalist7} & \multicolumn{1}{c|}{} \\
    \hline
    \multirow{1}{*}{019397d5} & \multirow{1}{*}{\textbf{24.89}} & \multirow{1}{*}{7.42} & \multirow{1}{*}{9.26} & \multirow{1}{*}{16.77} & \multirow{1}{*}{2.23} & \multirow{1}{*}{25.07}\\
    % & & & & & &\\
    \hline
    \multirow{1}{*}{3a099450} & \multirow{1}{*}{\textbf{42.82}} & \multirow{1}{*}{12.33} & \multirow{1}{*}{14.29} & \multirow{1}{*}{25.86} & \multirow{1}{*}{3.22} & \multirow{1}{*}{41.52}\\
    % & & & & & &\\
    \hline
    \multirow{1}{*}{4b4742a8} & \multirow{1}{*}{\textbf{22.1}} & \multirow{1}{*}{6.64} & \multirow{1}{*}{8.27} & \multirow{1}{*}{15.45} & \multirow{1}{*}{2.08} & \multirow{1}{*}{21.99}\\
    % & & & & & &\\
    \hline
    \multicolumn{1}{|c}{} & \multicolumn{5}{c}{Kodak} & \multicolumn{1}{c|}{} \\
    \hline
    \multirow{1}{*}{kodim01} & \multirow{1}{*}{22.75} & \multirow{1}{*}{7.48} & \multirow{1}{*}{11.03} & \multirow{1}{*}{20.38} & \multirow{1}{*}{2.39} & \multirow{1}{*}{\textbf{24.09}}\\
    % & & & & & &\\
    \hline
    \multirow{1}{*}{kodim02} & \multirow{1}{*}{\textbf{268.61}} & \multirow{1}{*}{87.54} & \multirow{1}{*}{82.28} & \multirow{1}{*}{157.16} & \multirow{1}{*}{20.19} & \multirow{1}{*}{242.13}\\
    % & & & & & &\\
    \hline
    \multirow{1}{*}{kodim03} & \multirow{1}{*}{\textbf{77.82}} & \multirow{1}{*}{22.48} & \multirow{1}{*}{25.71} & \multirow{1}{*}{47.38} & \multirow{1}{*}{6} & \multirow{1}{*}{75.91}\\
    % & & & & & &\\
    \hline
  \end{tabular}

  \caption{Compression ratio comparison for four datasets. Only for one image does WebP perform better than the  proposed algorithm.}
  \label{tab:table1}  
\end{table}

\section{Analytical Background}
\label{sec:bckgrnd}
In this section, we discuss statistical analysis of images and learned multi-level dictionaries that motivated this work and guided the design of the proposed approach. This analysis also provides various insights into the workings of the proposed approach and its performance.

Natural images and other types of images exhibit local similarity or smoothness. They are usually modeled as a Markov Random Field (MRF) due to similarity among neighboring pixels \cite{098rangarajanChellappa}. This means that for a pixel we expect the neighborhood pixels also to have similar values. When we quantize images to the level where each pixel is represented by a bit, wherever pixels in the RGB image have similar values, the neighboring pixels in the binary image would take the same value - either $0$ or $1$. Wherever there is an edge or other significant fluctuation of intensity in the RGB image, the binary values would exhibit a variation. This can be observed in the statistics that we have extracted.

Another consequence of the MRF nature of natural images is that, when we come to patches of size $n = 8,\,16$, though the number of possible patches is huge ($2^{64}$ and $2^{256}$, respectively), we see only a tiny fraction of patches in the binarized version of naturally occurring images. Below we give our observations on the statistics of the patterns of different sizes.

We also show that the probabilities of the top patterns (those having the highest frequency of occurrence) converge as we see more and more patches. This gives a lower bound on the amount of training data needed. Though in most cases, the amount of data needed for convergence is low, we continue the training to estimate the distribution of the patterns. 

\subsection{Bounded Dictionaries}
\label{subsec:bdddict}
In this subsection, we empirically establish that though a large number of unique patterns, for different block-sizes, may be observed over a collection of images, only a small number of those patterns occur most frequently and carry most of the frequency mass. This allows us to maintain manageable dictionary sizes even when a large number of patterns are observed and an even much larger number of patterns are possible, as in the case of patches of size $n = 8,\,16$. As further the patterns from these dictionaries are used to construct the Canonical Huffman codes (CHCs) for the patterns for a given block-size in an image, the bounded dictionary sizes ensure that such CHCs are shorter and are constructed with low computational cost.

\subsubsection{Dictionary for $2 \times 2$ patterns}
\begin{figure}[H]
  \centering
  \includegraphics[scale=0.4]{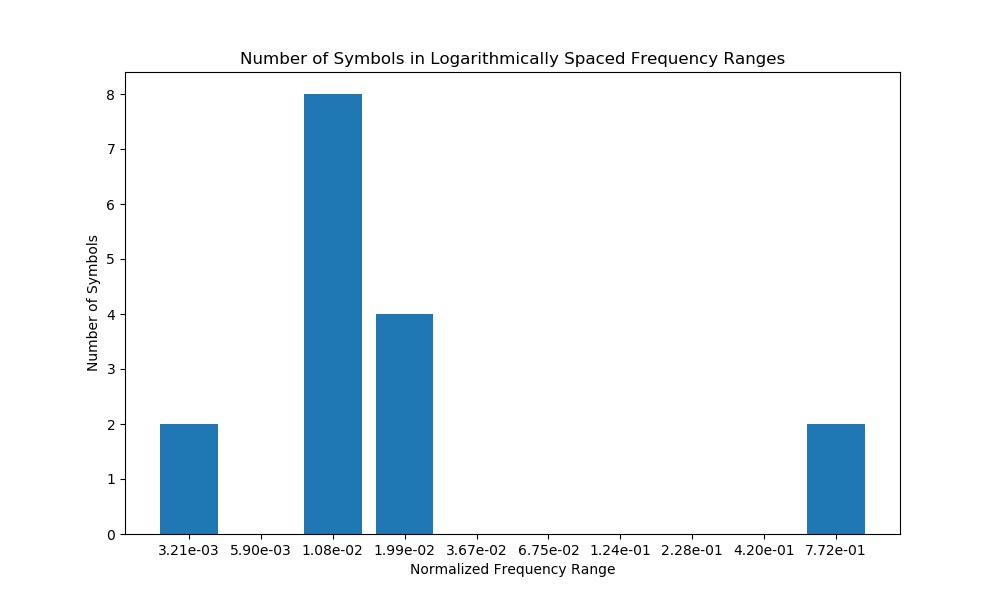}
  \caption{Pattern frequencies for $2 \times 2$ in a log-linear scale. Vertical bars represent the count of symbols lying in the particular frequency range. The frequency corresponding to the center of the interval is indicated along the X-axis.}
  \label{fig:2}
\end{figure}

There are only $16$ patterns of size $2 \times 2$ and as expected the frequency of the all zero and all ones patterns are both very high compared with the frequencies of other patterns. Figure~\ref{fig:2} shows the frequency plot obtained from the combined dataset (D1 and SAM).

The X-axis shows normalized frequency intervals (ratio of frequency intervals and the total number of patches used in training) in log scale and the Y-axis gives the number of symbols lying in each interval. It is observed that two symbols lie at the extreme right interval indicating that they both are of high frequency. These are the all zero and all one patterns. Similarly, we observe two patterns lying in the lowest frequency range and these correspond to symbols $6$ or $(0110)_2$ and $9$ or $(1001)_2$. In these two patterns, the binary value of the pixel changes with the highest spatial frequency.

\subsubsection{Dictionary for $4 \times 4$ patterns}
\begin{figure}[H]
  \centering
  \begin{subfigure}[b]{0.5\textwidth}
    \includegraphics[width=\textwidth]{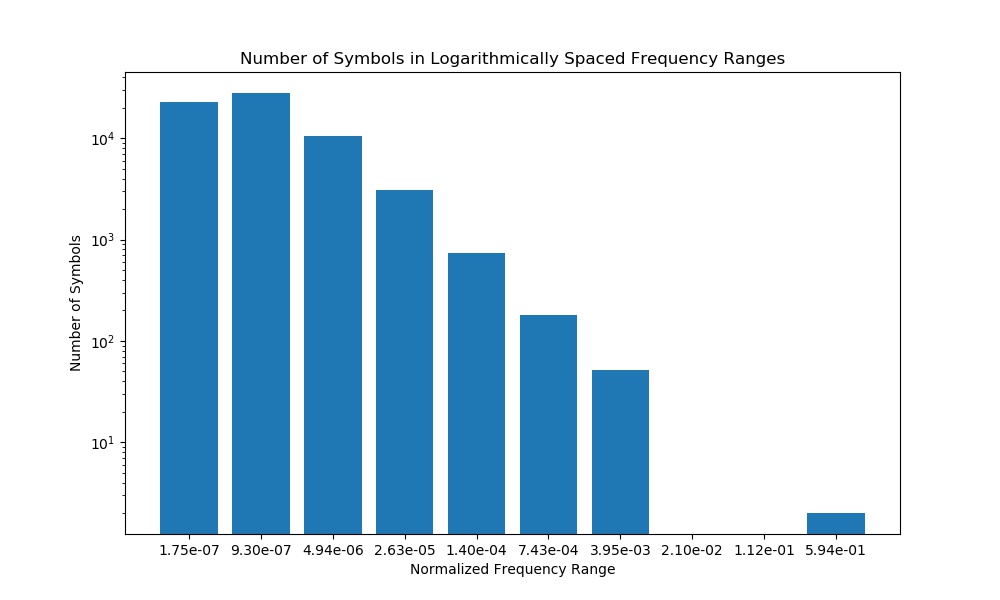}
    \caption{Number of patterns in normalized frequency intervals in log-log scale.}
    \label{fig:3a}
  \end{subfigure}
  \begin{subfigure}[b]{0.45\textwidth}
    \includegraphics[width=\textwidth,height=0.748\textwidth]{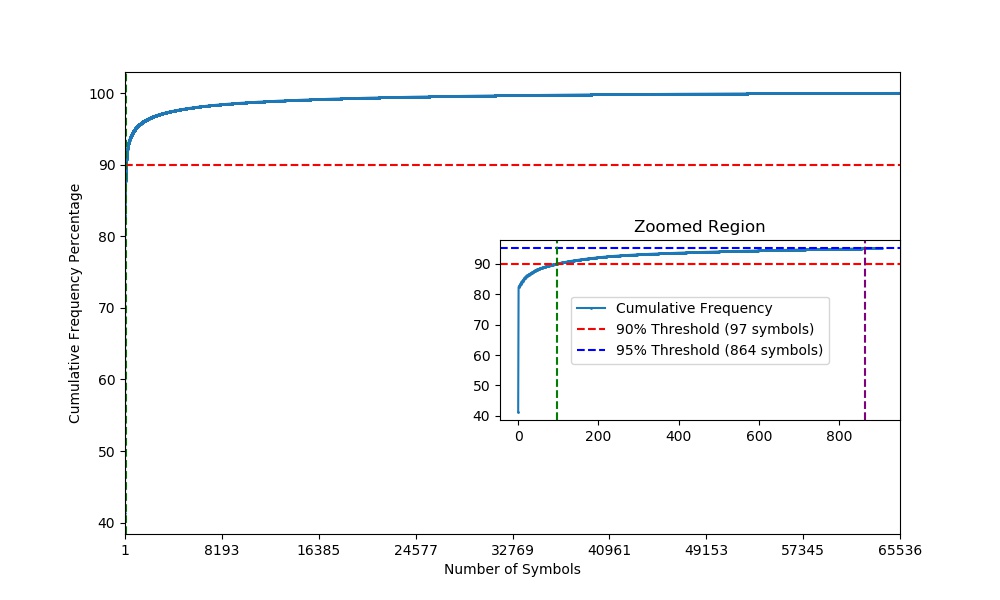}
    \caption{Cumulative sum plot.}
    \label{fig:3b}
  \end{subfigure}
  \caption{Statistics for $4 \times 4$ patterns.}
\end{figure}

For $4\times 4$ patterns, there are $65,536$ possible patterns and during dictionary learning, we encounter all those in the training data. The number of patterns (log scale) in each frequency interval (selected in a logarithmic manner) is shown in Figure~\ref{fig:3a} and the cumulative sum of frequencies is shown in Figure~\ref{fig:3b}.

From Figure~\ref{fig:3a}, it is seen that similar to the case of $2 \times 2$, two patterns occur with high frequencies. These are again the all zero and all one patterns. Many of the patterns cluster at lower frequency intervals showing that majority of the patterns are relatively rare. This trend is seen in Figure~\ref{fig:3b} too. It shows the pattern count along the X-axis and the cumulative frequency as a percentage along the Y-axis. The inset zoom on the initial region shows that only $97$ patterns contribute towards $90\%$ of the total frequency, and that $95\%$ of the total frequency is accounted for by only $864$ patterns. The rest of the $64,672$ patterns contribute only towards $5\%$ of the total frequency.

\subsubsection{Dictionary for $8 \times 8$ patterns}
\begin{figure}[H]
  \centering
  \begin{subfigure}[b]{0.5\textwidth}
    \includegraphics[width=\textwidth]{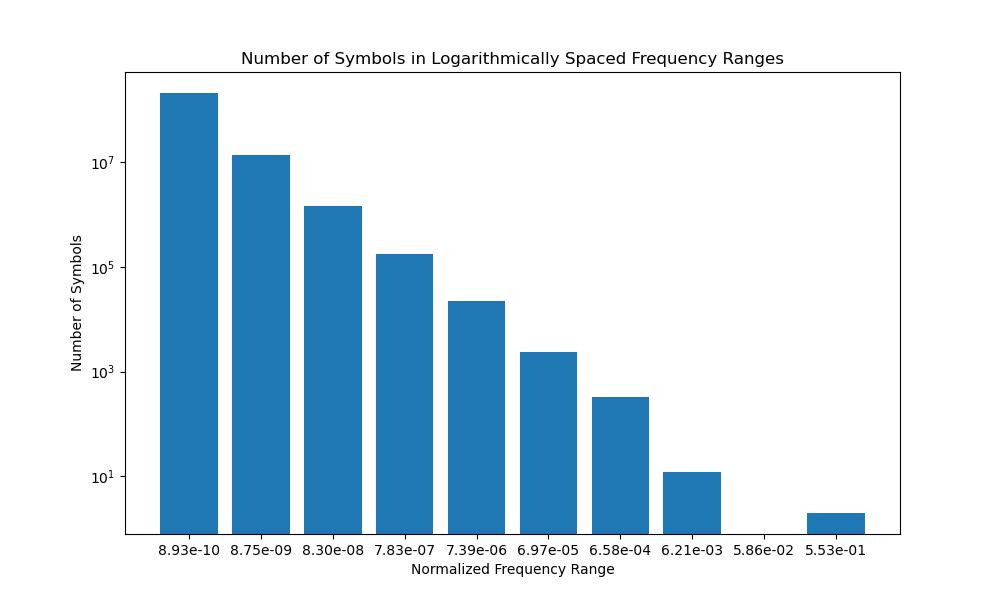}
    \caption{Number of patterns in normalized frequency intervals in log-log scale.}
    \label{fig:4a}
  \end{subfigure}
  \begin{subfigure}[b]{0.45\textwidth}
    \includegraphics[width=\textwidth,height=0.748\textwidth]{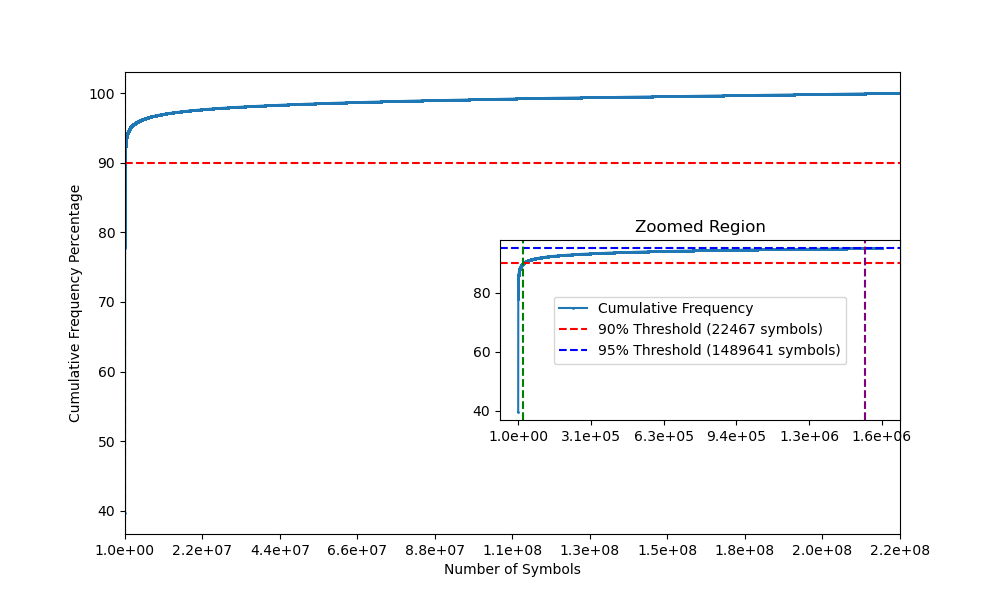}
    \caption{Cumulative sum plot.}
    \label{fig:4b}
  \end{subfigure}
  \caption{Statistics for $8 \times 8$ patterns.}
\end{figure}

Unlike in $2\times 2$ and $4 \times 4$, the number of patterns in this case is huge. However, the log-log frequency histogram plot in Figure~\ref{fig:4a} shows that most of the patterns are rare. The number of patterns in the lower frequency intervals are much higher than those at the high frequency ranges. The number of patterns occurring with unit frequency is $2,591,241,520$ and those with frequency two is $105,638,640$. It may be noted that we have discarded the unit frequency patterns while plotting these two graphs. It is seen from Figure~\ref{fig:4b} that from among roughly $2\times 10^8$ number of patterns, only around $20,000$ patterns contribute towards $90\%$ of the total frequency. This validates our argument that relatively only a few high frequency patterns come together to form an image. Further, as the probability of unit frequency patterns is close to zero, we do not consider them in creating the dictionary. 
   
\subsubsection{Dictionary for $16 \times 16$ patterns}
\begin{figure}[H]
  \centering
  \begin{subfigure}[b]{0.5\textwidth}
    \includegraphics[width=\textwidth]{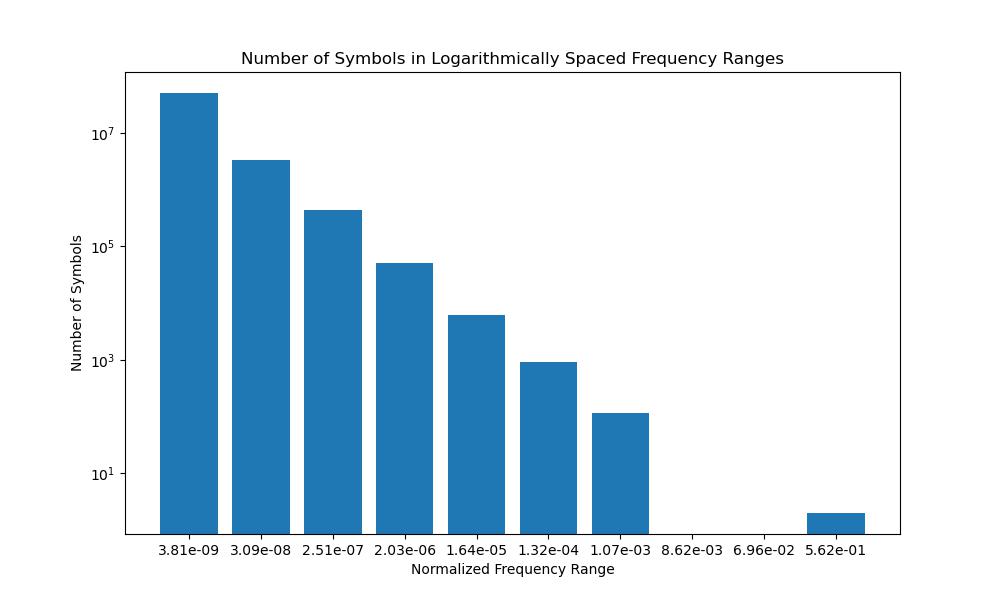}
    \caption{Number of patterns in normalized frequency intervals in log-log scale.}
    \label{fig:5a}
  \end{subfigure}
  \begin{subfigure}[b]{0.45\textwidth}
    \includegraphics[width=\textwidth,height=0.748\textwidth]{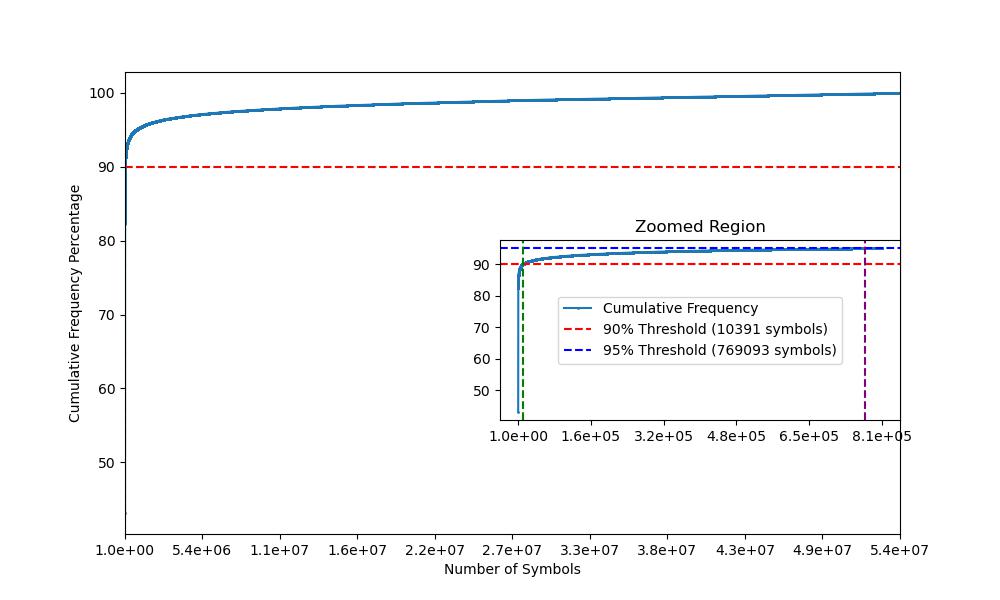}
    \caption{Cumulative sum plot}
    \label{fig:5b}
  \end{subfigure}
  \caption{Statistics for $16 \times 16$ patterns.}
\end{figure}

It is observed from Figures~\ref{fig:5a} and \ref{fig:5b}, that the statistics of $16 \times 16$ patterns also follow the trend of the statistics of $8 \times 8$ patterns. They differ only in the number of patterns seen, which is smaller compared to that of $8 \times 8$. The number of patterns with unit frequency is $2,248,110,545$ and the number of patterns with frequency $2$ is $26,607,341$. It may be noted that again we have discarded the unit frequency patterns while plotting both these graphs. It is seen from Figure~\ref{fig:5b} that from among roughly $5.5\times 10^7$ number of patterns, only around $10,000$ patterns contribute towards $90\%$ of the total frequency. Again, as the probability of unit frequency patterns is close to zero, we do not consider them in creating the dictionary. 

\subsection{Bounded Training}
\label{subsec:bddtrain}
In the last subsection, we established that even when a large number of patterns are possible for a given block-size, only a very small fraction of those patterns actually carry most of the frequency mass. In this subsection, we establish that for that small fraction of patterns, the probabilities converge with a low amount of training data. As these probabilities are used to construct corresponding CHCs for these patterns, we are assured that with a low amount of training and bounded dictionary size, we can construct CHCs that are robust against more training.

\subsubsection{Convergence of probabilities of $2 \times 2$ patterns}
\begin{figure}[H]
  \centering
  \begin{subfigure}[b]{0.3\textwidth}
    \includegraphics[scale=0.24]{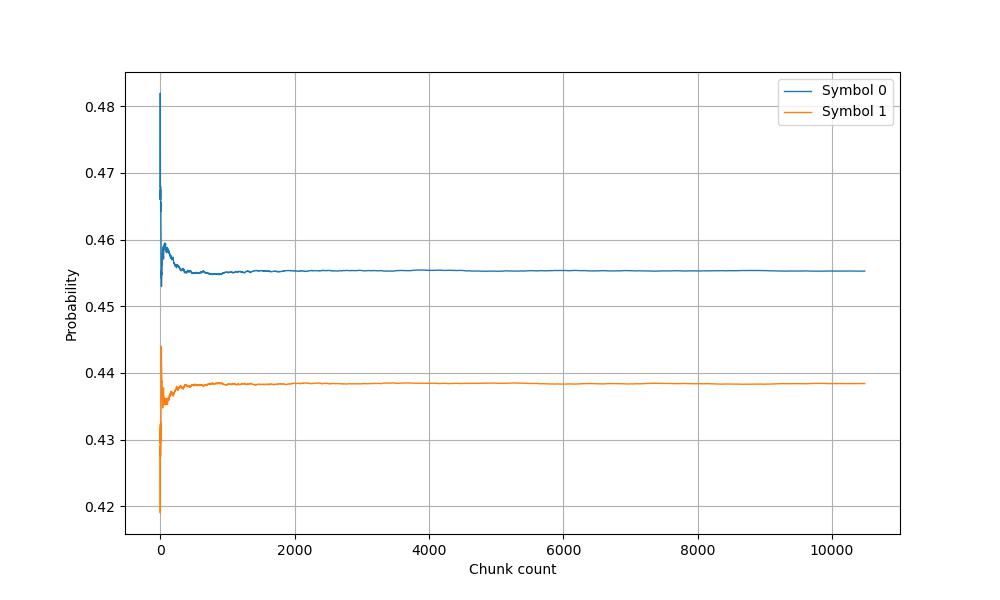}
    \caption{High frequency}
    \label{fig:p21}
  \end{subfigure}
  \begin{subfigure}[b]{0.3\textwidth}
    \includegraphics[scale=0.24]{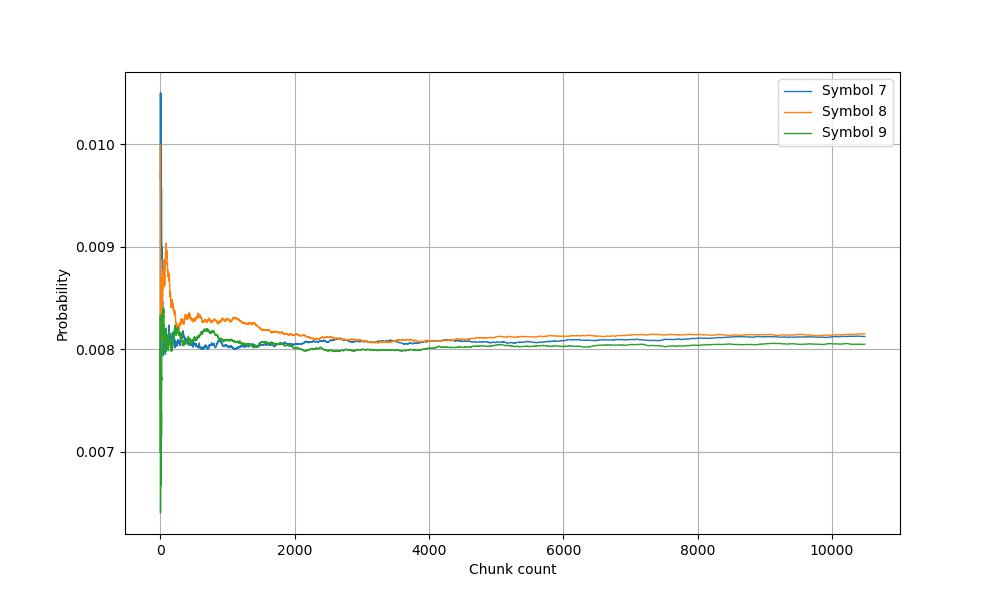}
    \caption{Mid frequency}
    \label{fig:p22}
  \end{subfigure}
  \begin{subfigure}[b]{0.3\textwidth}
    \includegraphics[scale=0.24]{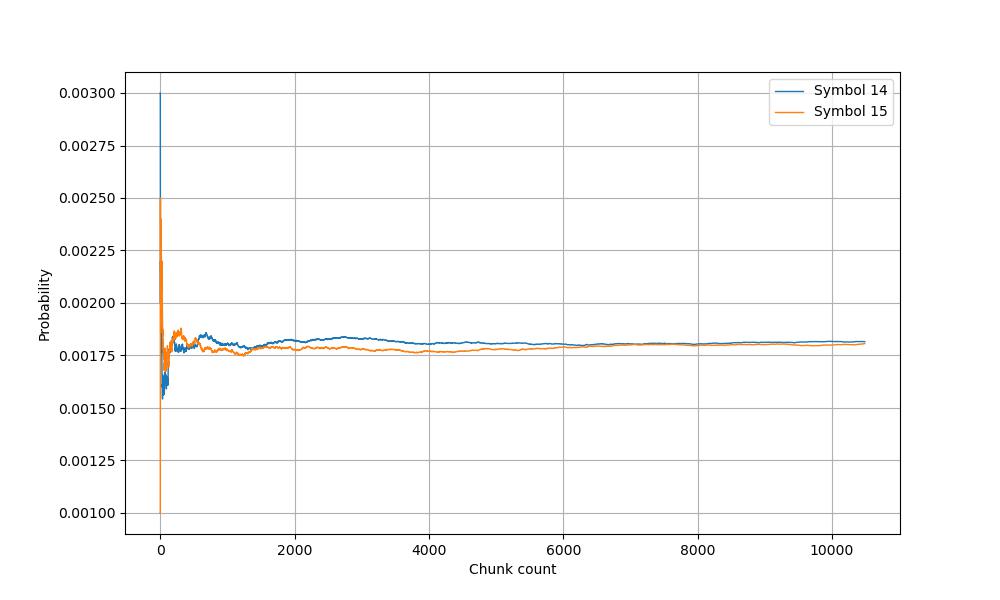}
    \caption{Low frequency}
    \label{fig:p23}
  \end{subfigure}
  \caption{Convergence of probabilities for $2\times 2$ patterns.}
\label{fig:PC2x2}
\end{figure}

Figure~\ref{fig:PC2x2} shows the probabilities of some of the patterns as a function of the number of patches observed. In order to see how the probabilities change with the number of observed patches, after reading and randomizing all possible patches from the images in a directory, the patches are split into chunks of size $1000$. The probabilities of the patterns are estimated chunk-wise. Figure~\ref{fig:PC2x2} shows this chunk count on the $X$-axis and the corresponding probabilities in the $Y$-axis. Figure~\ref{fig:p21} shows the convergence for the two most frequently occurring patterns ($0\texttt{xF}$ and $0\texttt{x0}$) and Figures~\ref{fig:p22} and \ref{fig:p23} give the same for three mid-frequency patterns and two of the lowest frequency patterns, respectively.

The most frequent patterns have probabilities converging to 0.45528 for pattern $0\texttt{x0}$ and 0.43841 for patterns $0xF$. Let $p_s[n]$ be the probability of pattern $s$ at the $n^{th}$ iteration. The condition for convergence is that
\begin{equation}
  \vert p_s[n] - p_s[n-k] \vert < \epsilon \quad \text{for} \quad k = 1,\,10,\,100,\,1000, \label{eq:conv}
  \end{equation}
where $\epsilon = 10^{-5}$, is true for all patterns. While the mid-frequency patterns converge to around $0.008$, the lowest frequency patterns converge to a probability around $0.0018$. Convergence according to Equation~\eqref{eq:conv} happened after processing $52.4$ MB of data. 

\subsubsection{Convergence of probabilities of $4 \times 4$ patterns}
\begin{figure}[H]
  \centering
  \begin{subfigure}[b]{0.3\textwidth}
    \includegraphics[scale=0.24]{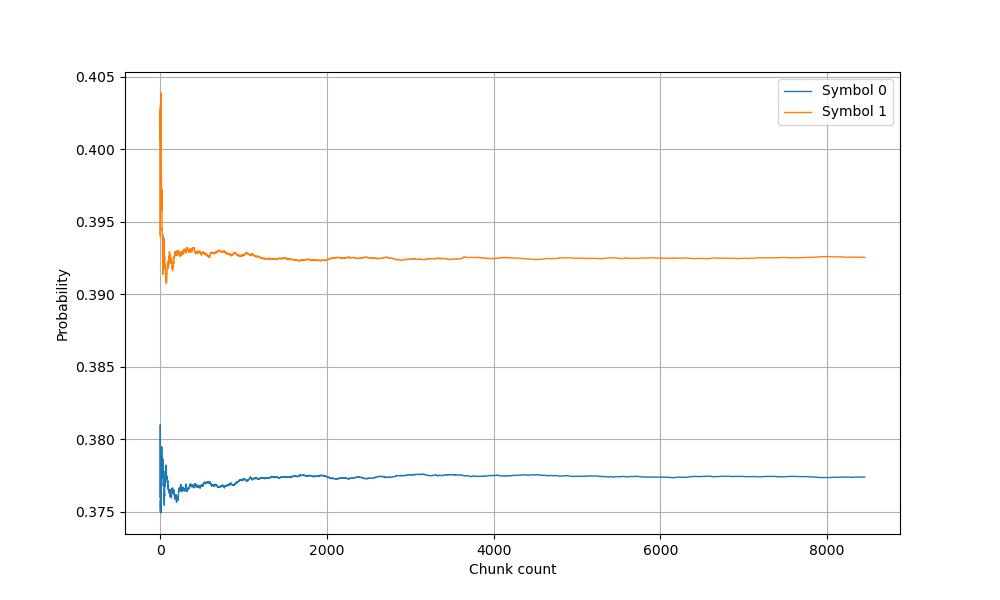}
    \caption{High frequency}
    \label{fig:p41}
  \end{subfigure}
  \begin{subfigure}[b]{0.3\textwidth}
    \includegraphics[scale=0.24]{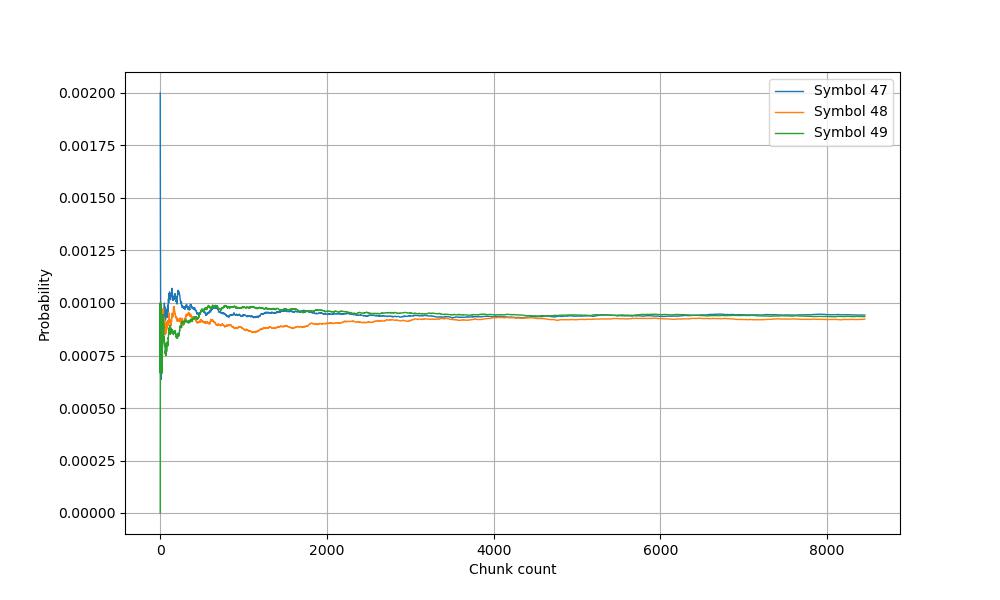}
    \caption{Mid frequency}
    \label{fig:p42}
  \end{subfigure}
  \begin{subfigure}[b]{0.3\textwidth}
    \includegraphics[scale=0.24]{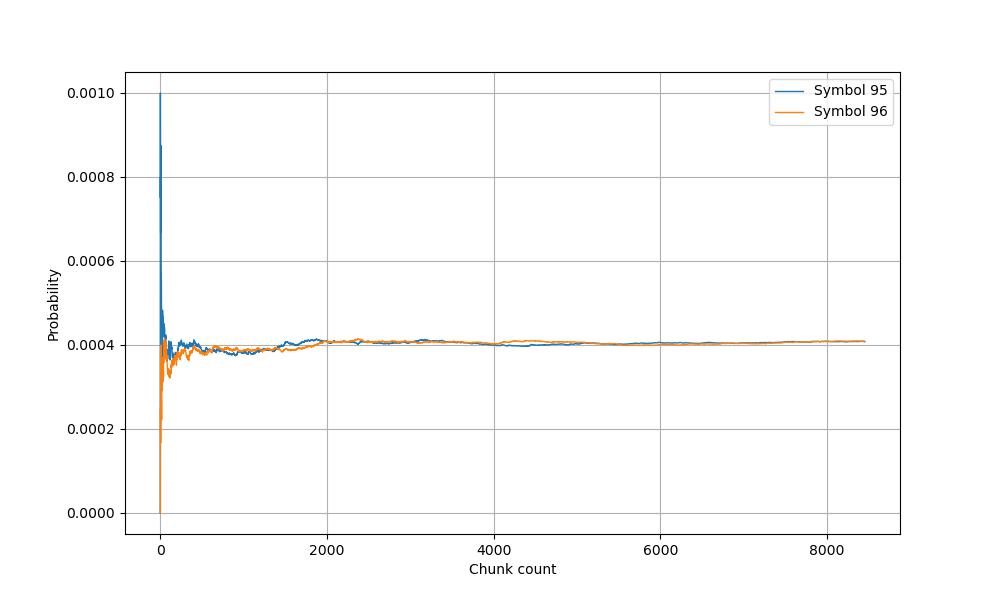}
    \caption{Low frequency}
    \label{fig:p43}
  \end{subfigure}
  \caption{Convergence of probabilities for $4\times 4$ patterns.}
\label{fig:PC4x4}
\end{figure}

In this case, convergence of probabilities is established for the top $97$ symbols that carry $90\%$ of the overall frequency mass of $4 \times 4$ patterns. The condition for convergence is same as for the $2 \times 2$ case. Figure~\ref{fig:PC4x4} shows the results for the patterns in three frequency ranges. Convergence is attained after processing $16.9$ MB of data. The top frequency patterns converge to probabilities, $0.37739$  and $0.39253$, respectively and the lowest frequency patterns converge to probabilities $0.0004079$, $0.0004090$, respectively.  

\subsubsection{Convergence of probabilities of $8 \times 8$ patterns}
\begin{figure}[H]
  \centering
  \begin{subfigure}[b]{0.3\textwidth}
    \includegraphics[scale=0.24]{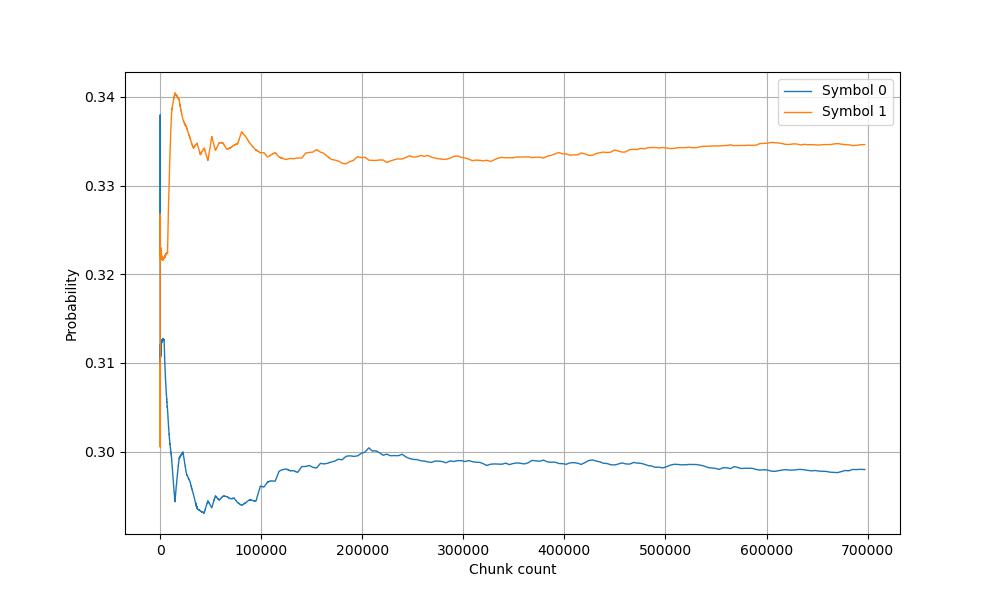}
    \caption{High frequency}
    \label{fig:p81}
  \end{subfigure}
  \begin{subfigure}[b]{0.3\textwidth}
    \includegraphics[scale=0.24]{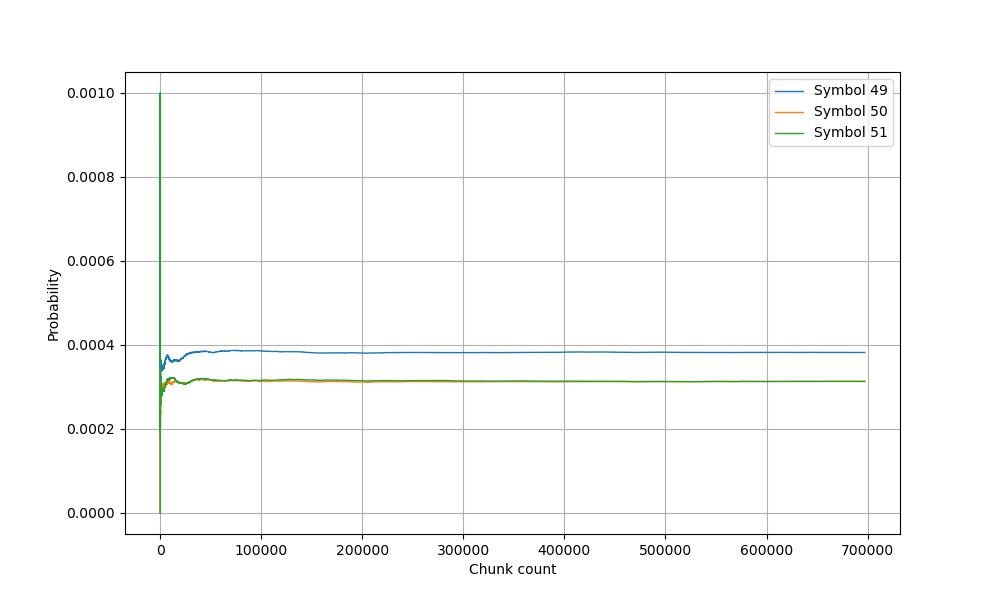}
    \caption{Mid frequency}
    \label{fig:p82}
  \end{subfigure}
  \begin{subfigure}[b]{0.3\textwidth}
    \includegraphics[scale=0.24]{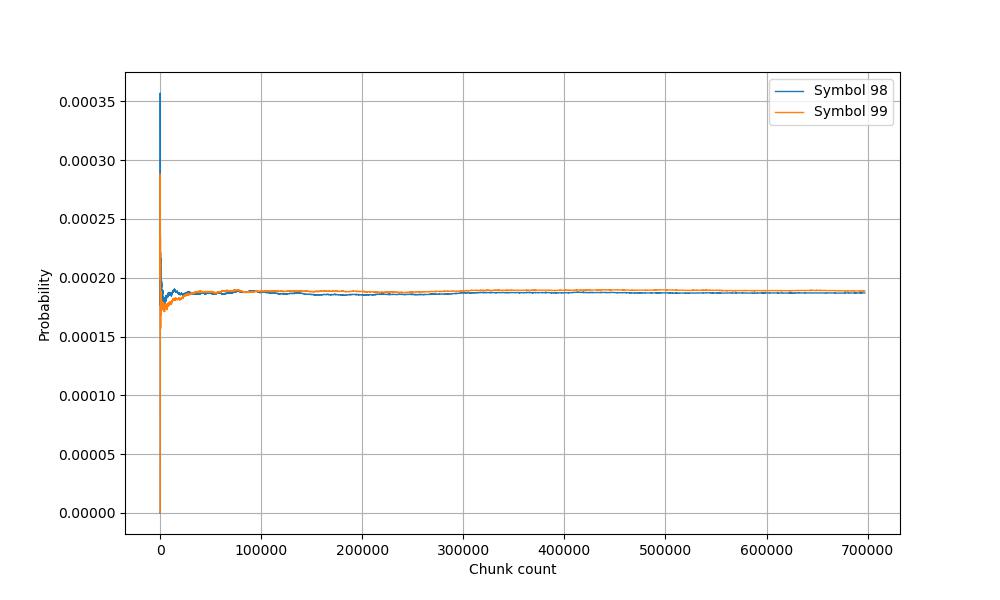}
    \caption{Low frequency}
    \label{fig:p83}
  \end{subfigure}
  \caption{Convergence of probabilities for $8\times 8$ patterns.}
\label{fig:PC8x8}
\end{figure}

Figure~\ref{fig:PC8x8} shows the way the probabilities converge for the $8 \times 8$ case. Here we consider only the top $100$ symbols for checking convergence. The condition for convergence is given in Equation~\eqref{eq:conv} with $\epsilon = 10^{-6}$. A total of $3748$ chunks are processed before convergence is attained. This is close to $5.6$ GB of data. The trend shown is similar to the previous two cases.

\subsubsection{Convergence of probabilities of $16 \times 16$ patterns}
Finally, in this case the convergence is established for the top $100$ patterns with the condition for convergence same as that for $8 \times 8$. For convergence, around $5.2$ GB of data is processed. The trend is similar to that of previous cases with the lower frequency symbol probabilities falling down to $8 \times 10^{-5}$.
\begin{figure}[H]
  \centering
  \begin{subfigure}[b]{0.3\textwidth}
    \includegraphics[scale=0.24]{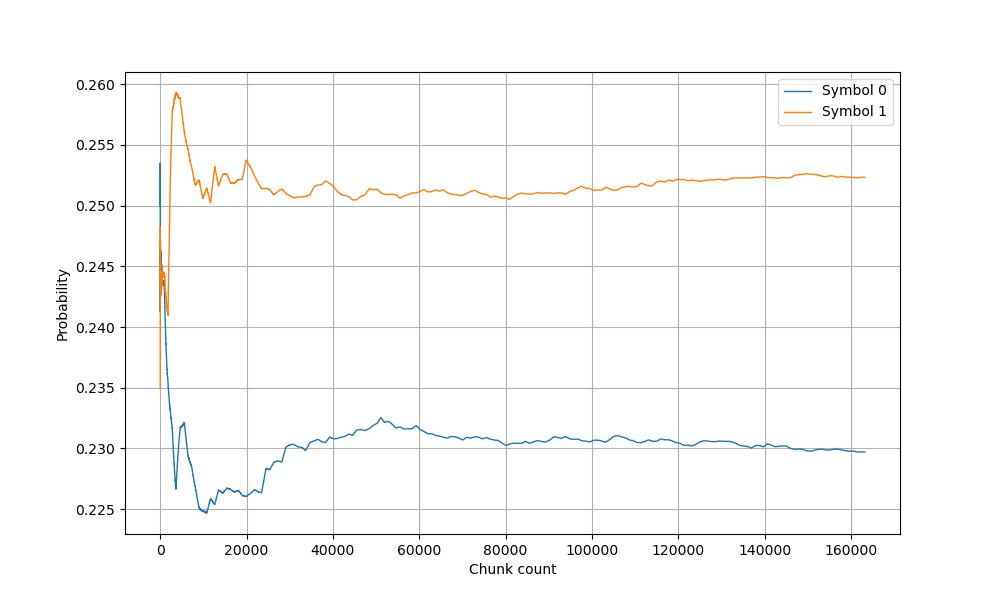}
    \caption{High frequency}
    \label{fig:p161}
  \end{subfigure}
  \begin{subfigure}[b]{0.3\textwidth}
    \includegraphics[scale=0.24]{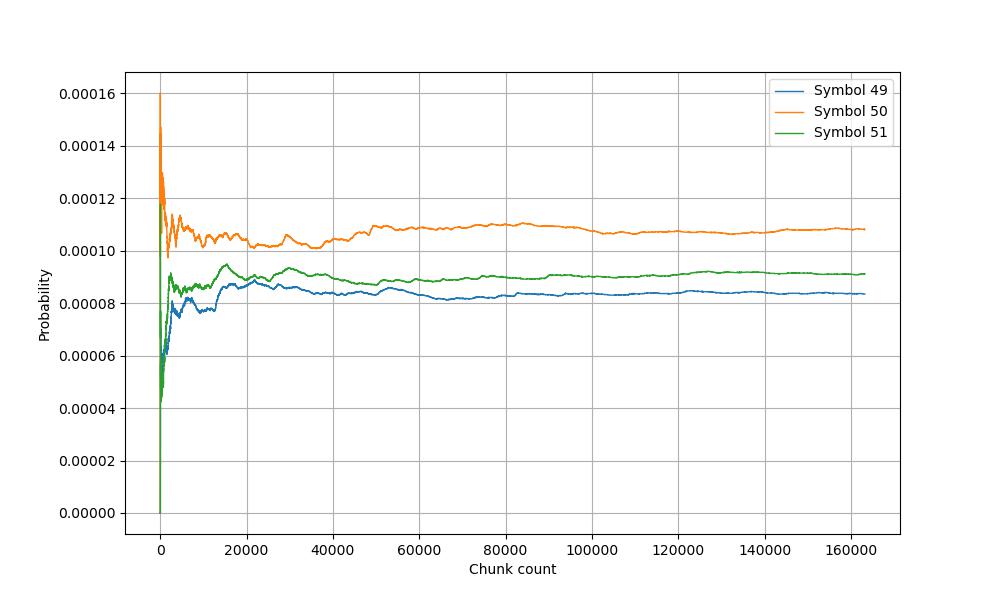}
    \caption{Mid frequency}
    \label{fig:p162}
  \end{subfigure}
  \begin{subfigure}[b]{0.3\textwidth}
    \includegraphics[scale=0.24]{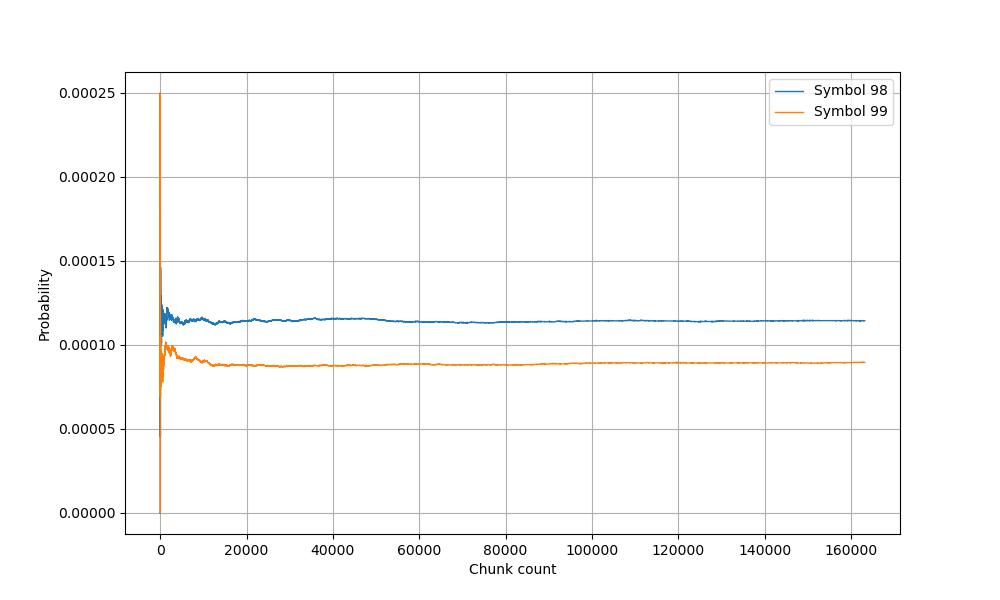}
    \caption{Low frequency}
    \label{fig:p163}
  \end{subfigure}
  \caption{Convergence of probabilities for $16 \times 16$ patterns.}
\label{fig:PC16x16}
\end{figure}

\section{Conclusions and Future Work}
\label{sec:concl}
We explore compression from a new perspective of the information contained at different quantization levels (color, intensity, and binary) and provide a scheme for lossless compression of binary images. The proposed multi-level dictionary based encoder outperforms the existing conventional as well as learning based image compression schemes by a large margin and has competitive performance with respect to a specialized scheme for compression of binary images.

Much work remains to be done. The current implementation of the proposed scheme only provides a proof of the concept. There exists various opportunities to optimize it to improve compression performance as well as computational requirements. In near future, we plan to work in these directions.

Also, we plan to formalize our work to extend the proposed  multi-level dictionary learning approach to develop schemes that exploit the patterns in intensity and color variations in images to construct lossless compression schemes for grayscale and color images.

%\newpage
%\bibliographystyle{IEEEtran}
%\begin{thebibliography}{1}

\newpage
\appendix
\renewcommand\thefigure{\thesection.\arabic{figure}}
\setcounter{figure}{0}
\renewcommand\thetable{\thesection.\arabic{table}}
\setcounter{table}{0}

%\subsection{Additional Results}
We give results on more images from four datasets. In each row, number in boldface corresponds to the scheme that provides the best compression ratio among all six schemes.

\begin{table}[H]
  \centering
  \begin{tabular}{|P{1.5cm}|P{1.5cm}|P{1.5cm}|P{1.5cm}|P{1.5cm}|P{1.5cm}|P{1.5cm}|}
    \hline
    \multicolumn{1}{|c|}{Image} & \multicolumn{1}{c|}{Proposed} & \multicolumn{1}{c|}{MGT} & \multicolumn{1}{c|}{PNG} & \multicolumn{1}{c|}{WebP} & \multicolumn{1}{c|}{JPEG-XL} &  \multicolumn{1}{c|}{JBIG2} \\
    \hline

    \multirow{1}{*}{n01518878} & \multirow{1}{*}{\textbf{67.76}} & \multirow{1}{*}{19.99} & \multirow{1}{*}{24.38} & \multirow{1}{*}{42.58}& \multirow{1}{*}{5.56}& \multirow{1}{*}{61.38}\\\hline
    \multirow{1}{*}{n02107142} & \multirow{1}{*}{\textbf{63.84}} & \multirow{1}{*}{21.96} & \multirow{1}{*}{27.16} & \multirow{1}{*}{48.41}& \multirow{1}{*}{6.93}& \multirow{1}{*}{61.71}\\\hline
    \multirow{1}{*}{n02454379} & \multirow{1}{*}{\textbf{55.67}} & \multirow{1}{*}{18.03} & \multirow{1}{*}{23.58} & \multirow{1}{*}{37.62}& \multirow{1}{*}{5.7} & \multirow{1}{*}{53.39}\\\hline
    \multirow{1}{*}{n02814860} & \multirow{1}{*}{\textbf{18.45}} & \multirow{1}{*}{6.65}  & \multirow{1}{*}{9.95}  & \multirow{1}{*}{16.70}& \multirow{1}{*}{2.5} & \multirow{1}{*}{18.59}\\\hline
    \multirow{1}{*}{n02910353.13} & \multirow{1}{*}{\textbf{128.63}} & \multirow{1}{*}{30.71} & \multirow{1}{*}{32.45} & \multirow{1}{*}{52.92}& \multirow{1}{*}{6.24}& \multirow{1}{*}{86.48}\\\hline
    \multirow{1}{*}{n02910353.15} & \multirow{1}{*}{22.69}           & \multirow{1}{*}{8.31}  & \multirow{1}{*}{12.4}  & \multirow{1}{*}{20.84}& \multirow{1}{*}{3.14}& \multirow{1}{*}{\textbf{23.33}}\\\hline
    \multirow{1}{*}{n02910353.58} & \multirow{1}{*}{\textbf{162.95}}& \multirow{1}{*}{54.10} & \multirow{1}{*}{50.09} & \multirow{1}{*}{99.84}& \multirow{1}{*}{13.21} & \multirow{1}{*}{145.24}\\\hline
    \multirow{1}{*}{n03196217} & \multirow{1}{*}{\textbf{337.88}}& \multirow{1}{*}{93.56} & \multirow{1}{*}{62.93}& \multirow{1}{*}{197.47}& \multirow{1}{*}{18.88} & \multirow{1}{*}{298.14}\\\hline
    \multirow{1}{*}{n03796401} & \multirow{1}{*}{122.99}         & \multirow{1}{*}{43.04} & \multirow{1}{*}{33.6}  & \multirow{1}{*}{86.34}& \multirow{1}{*}{10.13} & \multirow{1}{*}{\textbf{125.39}}\\\hline
    \multirow{1}{*}{n06359193} & \multirow{1}{*}{31.55}          & \multirow{1}{*}{10.51} & \multirow{1}{*}{15.72} & \multirow{1}{*}{31.68}& \multirow{1}{*}{3.82}  & \multirow{1}{*}{\textbf{32.25}}\\\hline
    \hline
  \end{tabular}
  \caption{Compression ratio comparison for images from ImageNet dataset.}
  \label{tab:table2}  
\end{table}

\begin{table}[H]
  \centering
  \begin{tabular}{|P{1.5cm}|P{1.5cm}|P{1.5cm}|P{1.5cm}|P{1.5cm}|P{1.5cm}|P{1.5cm}|}
    \hline
    \multicolumn{1}{|c|}{Image} & \multicolumn{1}{c|}{Proposed} & \multicolumn{1}{c|}{MGT} & \multicolumn{1}{c|}{PNG} & \multicolumn{1}{c|}{WebP} & \multicolumn{1}{c|}{JPEG-XL} &  \multicolumn{1}{c|}{JBIG2} \\
    \hline

    \multirow{1}{*}{sa\_6366087} & \multirow{1}{*}{28.4} & \multirow{1}{*}{9.43} & \multirow{1}{*}{13.33}& \multirow{1}{*}{22.89}& \multirow{1}{*}{3.47}&  \multirow{1}{*}{\textbf{31.75}} \\\hline
    \multirow{1}{*}{sa\_6367580} & \multirow{1}{*}{31.6} & \multirow{1}{*}{9.96} & \multirow{1}{*}{16.50}& \multirow{1}{*}{27.87}& \multirow{1}{*}{4.91} & \multirow{1}{*}{\textbf{39.77}}\\ \hline
    \multirow{1}{*}{sa\_6367649} & \multirow{1}{*}{40.66} & \multirow{1}{*}{13.87}& \multirow{1}{*}{16.76}& \multirow{1}{*}{32.27}& \multirow{1}{*}{4.69} & \multirow{1}{*}{\textbf{45.77}}\\ \hline
    \multirow{1}{*}{sa\_6367685} & \multirow{1}{*}{149.89}& \multirow{1}{*}{47.01}& \multirow{1}{*}{36.86}&\multirow{1}{*}{100.61}& \multirow{1}{*}{10.14}& \multirow{1}{*}{\textbf{168.2}}\\\hline
    \multirow{1}{*}{sa\_6369140} & \multirow{1}{*}{45.16} & \multirow{1}{*}{13.33}& \multirow{1}{*}{18.56}& \multirow{1}{*}{32.19}& \multirow{1}{*}{4.45} & \multirow{1}{*}{\textbf{51.69}}\\ \hline
    \multirow{1}{*}{sa\_6369201} & \multirow{1}{*}{44.53} & \multirow{1}{*}{13.68}& \multirow{1}{*}{18.39}& \multirow{1}{*}{32.39}& \multirow{1}{*}{4.50} & \multirow{1}{*}{\textbf{48.43}}\\ \hline
    \multirow{1}{*}{sa\_6372170} & \multirow{1}{*}{49.71} & \multirow{1}{*}{16.01}& \multirow{1}{*}{22.30}& \multirow{1}{*}{39.25}& \multirow{1}{*}{5.86} & \multirow{1}{*}{\textbf{53.82}}\\ \hline
    \multirow{1}{*}{sa\_6372464} & \multirow{1}{*}{141.03}& \multirow{1}{*}{42.93}& \multirow{1}{*}{50.09}& \multirow{1}{*}{86.48} &\multirow{1}{*}{11.46} &\multirow{1}{*}{\textbf{153.24}}\\\hline
    \multirow{1}{*}{sa\_6375426} & \multirow{1}{*}{87.57} & \multirow{1}{*}{30.18}& \multirow{1}{*}{34.68}& \multirow{1}{*}{69.05}& \multirow{1}{*}{10.15} &\multirow{1}{*}{\textbf{98.84}}\\ \hline
    \multirow{1}{*}{sa\_6375657} & \multirow{1}{*}{166.98}& \multirow{1}{*}{63.47}& \multirow{1}{*}{67.29}&\multirow{1}{*}{144.86}& \multirow{1}{*}{18.16} &\multirow{1}{*}{\textbf{200.96}}\\\hline
  \end{tabular}
  \caption{Compression ratio comparison for images from SAM dataset.}
  \label{tab:table3}  
\end{table}

\begin{table}[H]
  \centering
  \begin{tabular}{|P{1.5cm}|P{1.5cm}|P{1.5cm}|P{1.5cm}|P{1.5cm}|P{1.5cm}|P{1.5cm}|}
    \hline
    \multicolumn{1}{|c|}{Image} & \multicolumn{1}{c|}{Proposed} & \multicolumn{1}{c|}{MGT} & \multicolumn{1}{c|}{PNG} & \multicolumn{1}{c|}{WebP} & \multicolumn{1}{c|}{JPEG-XL} &  \multicolumn{1}{c|}{JBIG2} \\
    \hline

    \multirow{1}{*}{019397d5} & \multirow{1}{*}{\textbf{25.12}} & \multirow{1}{*}{7.42} &\multirow{1}{*}{9.26} &\multirow{1}{*}{16.77} &\multirow{1}{*}{2.23}&\multirow{1}{*}{25.07}\\ \hline
    \multirow{1}{*}{07515e5e} & \multirow{1}{*}{\textbf{35.47}} & \multirow{1}{*}{10.41}&\multirow{1}{*}{12.08}&\multirow{1}{*}{22.12} &\multirow{1}{*}{2.89}&\multirow{1}{*}{34.35}\\\hline  
    \multirow{1}{*}{3a099450} & \multirow{1}{*}{\textbf{42.82}} & \multirow{1}{*}{12.33}&\multirow{1}{*}{14.29}&\multirow{1}{*}{25.86} &\multirow{1}{*}{3.22}&\multirow{1}{*}{41.52}\\\hline
    \multirow{1}{*}{4b4742a8} & \multirow{1}{*}{\textbf{22.1}} &  \multirow{1}{*}{6.64} &\multirow{1}{*}{8.27} &\multirow{1}{*}{15.45} &\multirow{1}{*}{2.08}&\multirow{1}{*}{21.98}\\ \hline
    \multirow{1}{*}{4fc46e54} & \multirow{1}{*}{\textbf{107.17}} &\multirow{1}{*}{25.96}&\multirow{1}{*}{38.9} &\multirow{1}{*}{68.15} &\multirow{1}{*}{5.95}&\multirow{1}{*}{97.91}\\\hline
    \multirow{1}{*}{63af9699} & \multirow{1}{*}{15.58} & \multirow{1}{*}{5.22}          &\multirow{1}{*}{7.32} &\multirow{1}{*}{12.81} &\multirow{1}{*}{1.98}& \multirow{1}{*}{\textbf{16.12}}\\ \hline
    \multirow{1}{*}{662fb22c} & \multirow{1}{*}{9.81} & \multirow{1}{*}{3.52}           &\multirow{1}{*}{5.91} &\multirow{1}{*}{10.15} &\multirow{1}{*}{1.51}& \multirow{1}{*}{\textbf{10.31}}\\ \hline
    \multirow{1}{*}{6dcb6418} & \multirow{1}{*}{12.12} & \multirow{1}{*}{4.16}          &\multirow{1}{*}{6.43} &\multirow{1}{*}{10.90} &\multirow{1}{*}{1.59}& \multirow{1}{*}{\textbf{12.75}}\\ \hline
    \multirow{1}{*}{70dbd589} & \multirow{1}{*}{\textbf{202.62}} & \multirow{1}{*}{64.6}&\multirow{1}{*}{49.66}&\multirow{1}{*}{114.75}&\multirow{1}{*}{14.58}&\multirow{1}{*}{197.58}\\\hline
    \multirow{1}{*}{79b7c4cb} & \multirow{1}{*}{\textbf{79.31}}  &\multirow{1}{*}{23.22}&\multirow{1}{*}{25.09}&\multirow{1}{*}{48.05} &\multirow{1}{*}{6.38}&\multirow{1}{*}{76.62}\\\hline
  \end{tabular}                                                                                                                                                   
  \caption{Compression ratio comparison for images from iNaturalist dataset.}                             
  \label{tab:table4}                                                                                                                                                       
\end{table}

\begin{table}[H]
  \centering
  \begin{tabular}{|P{1.5cm}|P{1.5cm}|P{1.5cm}|P{1.5cm}|P{1.5cm}|P{1.5cm}|P{1.5cm}|}
    \hline
    \multicolumn{1}{|c|}{Image} & \multicolumn{1}{c|}{Proposed} & \multicolumn{1}{c|}{MGT} & \multicolumn{1}{c|}{PNG} & \multicolumn{1}{c|}{WebP} & \multicolumn{1}{c|}{JPEG-XL} &  \multicolumn{1}{c|}{JBIG2} \\
    \hline
    \multirow{1}{*}{kodim01}& \multirow{1}{*}{22.75}& \multirow{1}{*}{7.48}   & \multirow{1}{*}{11.03}& \multirow{1}{*}{20.38}&\multirow{1}{*}{2.39}&\multirow{1}{*}{\textbf{24.09}}\\\hline
    \multirow{1}{*}{kodim02}& \multirow{1}{*}{\textbf{268.61}}&\multirow{1}{*}{87.54}& \multirow{1}{*}{82.28}& \multirow{1}{*}{157.1}& \multirow{1}{*}{20.1}&\multirow{1}{*}{242.13}\\\hline
    \multirow{1}{*}{kodim03}& \multirow{1}{*}{\textbf{77.82}}& \multirow{1}{*}{22.48}& \multirow{1}{*}{25.71}& \multirow{1}{*}{47.38}& \multirow{1}{*}{6.00}& \multirow{1}{*}{75.91}\\\hline
    \multirow{1}{*}{kodim04}& \multirow{1}{*}{\textbf{55.40}}& \multirow{1}{*}{17.60}& \multirow{1}{*}{21.46}& \multirow{1}{*}{38.52}& \multirow{1}{*}{5.33}& \multirow{1}{*}{53.32}\\\hline
    \multirow{1}{*}{kodim05}& \multirow{1}{*}{30.18}& \multirow{1}{*}{9.31 }& \multirow{1}{*}{12.19}& \multirow{1}{*}{20.74}& \multirow{1}{*}{3.07}& \multirow{1}{*}{\textbf{31.69}}\\\hline
    \multirow{1}{*}{kodim06}& \multirow{1}{*}{39.10}& \multirow{1}{*}{13.02}& \multirow{1}{*}{22.00}& \multirow{1}{*}{37.93}& \multirow{1}{*}{4.62}& \multirow{1}{*}{\textbf{39.97}}\\\hline
    \multirow{1}{*}{kodim07}& \multirow{1}{*}{60.61}& \multirow{1}{*}{18.11}& \multirow{1}{*}{20.62}& \multirow{1}{*}{38.66}& \multirow{1}{*}{4.57}& \multirow{1}{*}{\textbf{61.32}}\\\hline
    \multirow{1}{*}{kodim08}& \multirow{1}{*}{36.88}& \multirow{1}{*}{11.99}& \multirow{1}{*}{12.69}& \multirow{1}{*}{26.98}& \multirow{1}{*}{3.71}& \multirow{1}{*}{\textbf{37.45}}\\\hline
    \multirow{1}{*}{kodim09}& \multirow{1}{*}{54.91}& \multirow{1}{*}{17.56}& \multirow{1}{*}{22.06}& \multirow{1}{*}{44.00}& \multirow{1}{*}{5.37}& \multirow{1}{*}{\textbf{55.57}}\\\hline
    \multirow{1}{*}{kodim10}& \multirow{1}{*}{\textbf{70.08}}& \multirow{1}{*}{21.20}& \multirow{1}{*}{23.35}& \multirow{1}{*}{42.62}& \multirow{1}{*}{5.93}& \multirow{1}{*}{68.21}\\\hline
    \multirow{1}{*}{kodim11}& \multirow{1}{*}{\textbf{48.96}}& \multirow{1}{*}{15.62}& \multirow{1}{*}{20.09}& \multirow{1}{*}{37.07}& \multirow{1}{*}{4.69}& \multirow{1}{*}{47.87}\\\hline
    \multirow{1}{*}{kodim12}& \multirow{1}{*}{\textbf{185.10}}&\multirow{1}{*}{56.14}& \multirow{1}{*}{49.37}& \multirow{1}{*}{109.78}&\multirow{1}{*}{13.27}&\multirow{1}{*}{166.12}\\\hline
    \multirow{1}{*}{kodim13}& \multirow{1}{*}{26.55}& \multirow{1}{*}{8.60 }& \multirow{1}{*}{14.50}& \multirow{1}{*}{24.04}& \multirow{1}{*}{3.05}& \multirow{1}{*}{\textbf{26.87}}\\\hline
    \multirow{1}{*}{kodim14}& \multirow{1}{*}{31.64}& \multirow{1}{*}{9.87 }& \multirow{1}{*}{14.74}& \multirow{1}{*}{24.51}& \multirow{1}{*}{3.14}& \multirow{1}{*}{\textbf{31.93}}\\\hline
    \multirow{1}{*}{kodim15}& \multirow{1}{*}{\textbf{182.00}}&\multirow{1}{*}{60.35}& \multirow{1}{*}{49.02}& \multirow{1}{*}{104.86}&\multirow{1}{*}{15.72}&\multirow{1}{*}{176.33}\\\hline
    \multirow{1}{*}{kodim16}& \multirow{1}{*}{40.80}& \multirow{1}{*}{13.35}& \multirow{1}{*}{22.81}& \multirow{1}{*}{38.18}& \multirow{1}{*}{4.65}& \multirow{1}{*}{\textbf{40.86}}\\\hline
    \multirow{1}{*}{kodim17}& \multirow{1}{*}{\textbf{56.03}}& \multirow{1}{*}{17.49}& \multirow{1}{*}{19.97}& \multirow{1}{*}{36.94}& \multirow{1}{*}{4.95}& \multirow{1}{*}{54.91}\\\hline
    \multirow{1}{*}{kodim18}& \multirow{1}{*}{\textbf{30.10}}& \multirow{1}{*}{9.70 }& \multirow{1}{*}{13.95}& \multirow{1}{*}{23.52}& \multirow{1}{*}{3.29}& \multirow{1}{*}{29.78}\\\hline
    \multirow{1}{*}{kodim19}& \multirow{1}{*}{72.47}& \multirow{1}{*}{23.21}& \multirow{1}{*}{23.50}& \multirow{1}{*}{51.50}& \multirow{1}{*}{7.62}& \multirow{1}{*}{\textbf{73.78}}\\\hline
    \multirow{1}{*}{kodim20}& \multirow{1}{*}{\textbf{198.11}}&\multirow{1}{*}{59.15}& \multirow{1}{*}{58.53}& \multirow{1}{*}{123.50}&\multirow{1}{*}{15.47}&\multirow{1}{*}{189.59}\\\hline
    \multirow{1}{*}{kodim21}& \multirow{1}{*}{42.56}& \multirow{1}{*}{13.74}& \multirow{1}{*}{21.20}& \multirow{1}{*}{35.96}& \multirow{1}{*}{4.46}& \multirow{1}{*}{\textbf{42.83}}\\\hline
    \multirow{1}{*}{kodim22}& \multirow{1}{*}{\textbf{48.47}}& \multirow{1}{*}{15.49}& \multirow{1}{*}{19.85}& \multirow{1}{*}{35.06}& \multirow{1}{*}{4.72}& \multirow{1}{*}{46.32}\\\hline
    \multirow{1}{*}{kodim23}& \multirow{1}{*}{\textbf{133.06}}&\multirow{1}{*}{41.50}& \multirow{1}{*}{39.61}& \multirow{1}{*}{76.15}& \multirow{1}{*}{11.06}&\multirow{1}{*}{126.84}\\\hline
    \multirow{1}{*}{kodim24}& \multirow{1}{*}{\textbf{60.88}}& \multirow{1}{*}{19.62}& \multirow{1}{*}{27.77}& \multirow{1}{*}{45.98}& \multirow{1}{*}{8.11}& \multirow{1}{*}{59.49}\\\hline
    
  \end{tabular}
  \caption{Compression ratio comparison for images from Kodak dataset.}
  \label{tab:table5}  
\end{table}

\end{document}